\documentclass{ar-1col}
\usepackage[comma]{natbib}
\usepackage{url}

\usepackage[T1]{fontenc}
\usepackage{amsmath}	
\usepackage{amssymb}	
\usepackage{enumitem}
\newcommand\aap{A\&A}                % Astronomy and Astrophysics
\newcommand\aj{AJ}                   % Astronomical Journal (the)
\newcommand\apj{ApJ}                 % Astrophysical Journal
\newcommand\apjl{ApJ}                % Astrophysical Journal, Letters
\newcommand\apjs{ApJS}               % Astrophysical Journal, Supplement
\newcommand\araa{ARA\&A}             % Annual Review of Astronomy and Astrophysics
\newcommand\mnras{MNRAS}             % Monthly Notices of the Royal Astronomical Society
\newcommand\nat{Nature}              % Nature
\newcommand\pasp{PASP}               % Publications of the Astronomical Society of the Pacific
\newcommand\rmxaa{Rev. Mex. Astron. Astrofis.} % Revista Mexicana de Astronomia y Astrofisica
\newcommand{\percmsq}           {{\rm cm}^{-2}}
\newcommand{\percc}           {{\rm cm}^{-3}}
\newcommand{\HI}              {H{\,\textsc{i}}}

\newcommand{\OVI}              {O{\,\textsc{vi}}}
\newcommand{\OVII}              {O{\,\textsc{vii}}}
\newcommand{\OVIII}              {O{\,\textsc{viii}}}
\newcommand{\K}            {\rm{K}}
\newcommand{\yr}            {\rm{yr}}
\newcommand{\Myr}            {\rm{Myr}}
\newcommand{\Gyr}            {\rm{Gyr}}
\newcommand{\Msun}            {\rm{M}_\odot}
\newcommand{\Mpc}            {\rm{Mpc}}
\newcommand{\cMpc}            {\,\rm{cMpc}}

\newcommand{\kpc}            {\,\rm{kpc}}

\newcommand{\pkpc}            {\,\rm{pkpc}}
\newcommand{\pc}            {\,\rm{pc}}

\newcommand{\perMpc}        {\,\rm{Mpc}^{-1}}

\jname{Annu. Rev. Astron. Astrophys.}
\jvol{AA}
\jyear{2023}
\doi{10.1146/annurev-astro-041923-043618}

\begin{document}

% Page header
\markboth{Crain \& van de Voort}{Simulations of the Galaxy Population}

% Title
\title{Hydrodynamical simulations of the galaxy population: enduring successes and outstanding challenges}

%Authors, affiliations address.
 \author{Robert A.\ Crain$^{1\star}$ and Freeke van de Voort$^{2\star}$
 \affil{$^1$Astrophysics Research Institute, Liverpool John Moores University, Liverpool, United Kingdom; email: r.a.crain@ljmu.ac.uk}
 \affil{$^2$Cardiff Hub for Astrophysics Research and Technology, School of Physics and Astronomy, Cardiff University, Cardiff, United Kingdom; email: freeke@astro.cf.ac.uk}
 \affil{$^\star$The authors contributed equally to this manuscript.} 
 }

%Abstract
\begin{abstract}
We review the progress in modelling the galaxy population in hydrodynamical simulations of the $\Lambda$CDM cosmogony. State-of-the-art simulations now broadly reproduce the observed spatial clustering of galaxies, the distributions of key characteristics such as mass,  size and star formation rate, and scaling relations connecting diverse properties to mass. Such improvements engender confidence in the insight drawn from simulations. Many important outcomes however, particularly the properties of circumgalactic gas, are sensitive to the details of the subgrid models used to approximate the macroscopic effects of unresolved physics, such as feedback processes. We compare the outcomes of leading simulation suites with observations and with each other, to identify the enduring successes they have cultivated and the outstanding challenges to be tackled with the next generation of models. Our key conclusions are:
\begin{itemize}[leftmargin=*]
    \item Realistic galaxies can be reproduced by calibrating the ill-constrained\\ parameters of subgrid feedback models. Feedback is dominated by stars\\ and by black holes in low mass and high mass galaxies, respectively.
    \item Adjusting or disabling the physical processes implemented in simulations\\ can elucidate their impact on observables, but outcomes can be degenerate.
    \item Similar galaxy populations can emerge in simulations with dissimilar\\ subgrid feedback implementations. However, these models  generally\\ predict markedly different gas flow rates into, and out of, galaxies and\\ their haloes. CGM observations are thus a promising means of breaking\\ this degeneracy and guiding the development of new feedback models.
\end{itemize}
\end{abstract}

%Keywords, etc.
\maketitle

%Table of Contents
\tableofcontents

\section{Introduction} \label{sec:intro}

The present-day galaxy population exhibits a remarkable diversity of characteristics, such as masses, star formation rates, morphologies, nuclear activity, and gas and dust content. The population's spatial distribution is highly-structured and heterogeneous: galaxies can be found in isolation in low density environments, or can comprise the population of rich clusters. Observations of the distant cosmos reveal that both the properties and spatial distribution of the galaxy population have evolved markedly over nearly 14 billion years of cosmic history. Reconciliation of this panoply of observed characteristics with a comprehensive theory of galaxy formation is a challenge at the frontier of the natural sciences. 

Owing to the non-linearity of the collapse, hierarchical assembly, and relaxation of protogalactic structure, and the complexity of the myriad physical processes that influence galaxies, direct numerical simulation of the evolution of representative cosmological volumes is in principle the most desirable method of approaching the challenge. In recent years, this approach has emerged as the foremost means of interpreting observations, particularly those from highly-multiplexed galaxy surveys, and of seeking a clearer understanding of the origin of galaxy properties. The widespread adoption of hydrodynamical simulations of representative cosmological volumes has been driven primarily by major improvements in their correspondence with the observed galaxy population, coupled with improved access to the simulation data, a development catalysed by the public release of data from major simulation campaigns \citep[e.g.][]{Nelson2015,Nelson2019,McAlpine2016, VillaescusaNavarro2023}. 

It is easy to now take for granted the availability of these versatile and realistic models. However, although simulations of representative volumes have yielded good agreement with observations of diffuse intergalactic gas (e.g.\ as traced by the Lyman-$\alpha$ forest) for over two decades \citep[][]{Theuns1998,Dave1999}, reproducing the properties and spatial distribution of galaxies and their star-forming gas reservoirs has proven a more stubborn challenge. For many years, simulated galaxies formed far too many stars, particularly at early cosmic epochs when the cosmic inflow rate is high, thus ending up too massive, too compact, and with too little angular momentum \citep[e.g.][]{Navarro1991,Navarro1995,SommerLarsen1999,Navarro2000}. Unsurprisingly then, they also exhibited unrealistic surface density and rotation velocity profiles \citep{Abadi2003a,Abadi2003b}. Simulations of representative cosmic volumes yielded galaxy stellar mass functions (GSMFs) with the wrong shape and normalisation, generally yielding too many galaxies at fixed stellar mass \citep[e.g.][]{Crain2009,Lackner2012,Khandai2015}. 
% Definitions
\begin{marginnote}[]
\entry{GSMF}{Galaxy stellar mass function}
\end{marginnote}

Mitigation of this `overcooling' of gas into stars, and the concomitant spurious transfer of angular momentum from gas to dark matter (DM), was demonstrated early in the history of hydrodynamical simulations of galaxies via the inclusion of gas heating mechanisms \citep[e.g.][]{Katz1991,Mihos1994,Navarro1997,Weil1998}. The inclusion of energetic `feedback' mechanisms, as a means to regulate gas cooling and star formation, has fostered major improvements in the realism of simulated galaxies, for example enabling the realisation of individual disc galaxies with encouragingly realistic surface density and rotation profiles \citep[e.g.][]{Governato2004,Okamoto2005,Governato2007,Guedes2011}.

However, the macroscopic efficiencies of feedback mechanisms are governed by microphysics acting on spatial scales several orders of magnitude below the resolution scale of galaxy population simulations \citep[e.g.][]{Orlando2005}, precluding their calculation \textit{ab initio}. These processes (and others) are therefore treated with simplified subgrid models (see Section \ref{sec:subgrid_methods}), which approximate the effects of unresolved processes and couple them to numerically-resolved scales, thereby producing an 'effective' model of galaxy formation. In the absence of authoritative empirical constraints on how microphysics influences macroscopic scales \citep[though see][]{Lopez2011,Rosen2014}, the subgrid implementations of feedback in popular simulation codes that have emerged are diverse, and can produce conspicuously dissimilar outcomes when applied to identical initial conditions \citep[see][and references therein]{Scannapieco2012}. The dramatic variation of outcomes that emerge from controlled suites of simulations in which (only) the subgrid implementation of feedback is changed, and/or their parameters are varied systematically over plausible ranges, highlights that the influence of feedback is the most important systematic uncertainty in galaxy formation modelling \citep{Oppenheimer2010,Schaye2010,Vogelsberger2013,Kim2014}. 

\begin{figure}[h]
\includegraphics[width=5in]{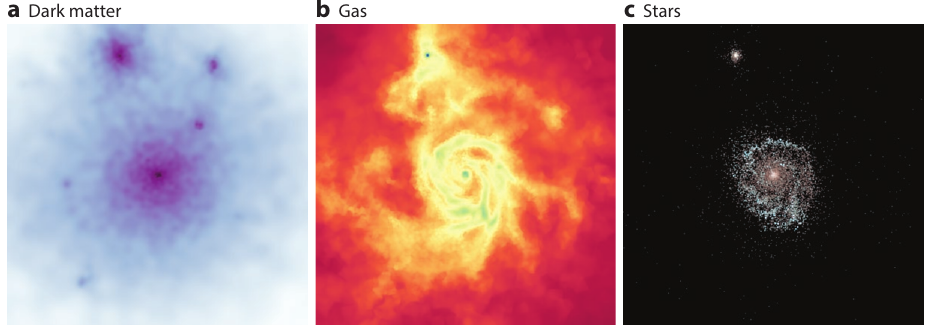}
\caption{An illustrative example of the level of detail with which modern cosmological simulations are able to reproduce the structure of galaxies similar in mass ($M_{halo}\sim 10^{12}~\Msun$) to the Milky Way, based on \texttt{IllustrisTNG}. From left to right, the panels show the dark matter surface density, the gas surface density, and a three colour image of the stellar surface brightness from the $U$, $B$ and $K$ bands. Each image has a 140~kpc field of view.  The central galaxy is gas-rich, star-forming and kinematically disc-dominated, with a significant bulge component. It is accompanied by a gas-rich satellite galaxy and dark matter substructure without an associated baryonic component. The gas density shows clumpy and filamentary structure not present in the dark matter and stars.}
\label{fig:MW}
\end{figure}

The most productive strategy that has emerged in response to this uncertainty is to calibrate the parameters of subgrid feedback models (see Section \ref{sec:subgrid_calibration}), with the aim of reproducing judiciously-chosen observable characteristics of the galaxy population. Clearly, the characteristics used for the calibration cannot be considered as predictions of the simulation, but other properties can be considered as outcomes stemming from the implemented physical processes, as long as they are not compromised by the simplifications of the subgrid modelling. This approach has been used, to varying degrees, by several flagship-scale simulations of the galaxy population, including \texttt{Illustris} \citep{Vogelsberger2014mn,Genel2014,Nelson2015}, \texttt{Magneticum}  \citep{Hirschmann2014}, \texttt{EAGLE} \citep{Schaye2015,Crain2015,McAlpine2016}, \texttt{BlueTides} \citep{Feng2016}, \texttt{Romulus} \citep{Tremmel2017}, \texttt{IllustrisTNG} \citep{Pillepich2018a,Nelson2018colour,Springel2018}, \texttt{FABLE}  \citep{Henden2018} and \texttt{SIMBA} \citep{Dave2019}. An example of the dark matter, gas, and stellar light distributions that emerge from simulating a Milky Way-mass dark matter halo with a modern galaxy formation model is shown in Figure~\ref{fig:MW}. The central and satellite galaxies visible have a clear gaseous and dark matter component as well. Further substructure is present in the dark matter and gas without a counterpart in either of the other components. 

% Definitions
\begin{marginnote}[]
\entry{IGM}{Intergalactic medium}
\end{marginnote}
The key advantage of the calibration approach is that it steers simulations toward the production of a broadly realistic simulated galaxy population. The galaxies interact self-consistently with the intergalactic medium (IGM), enabling the address of diverse lines of enquiry. Several of the aforementioned projects have demonstrated, for example, that it is possible to reproduce the present-day galaxy stellar mass function, from the mass scale of dwarf galaxies to that of central group galaxies, with an accuracy comparable to the systematic uncertainty on the observational measurement. A subset of the models also consider diagnostics inferred from X-ray observations of galaxy groups/clusters, in order to ensure reasonable reproduction of the properties of intragroup/intracluster gas. Simulations of realistic galaxies also enable additional model components to be `bolted on' either on-the-fly or in post-processing, for example to follow the formation and evolution of globular clusters \citep{Pfeffer2018} or to predict rates of gamma ray bursts \citep{Metha2020}. The realism and versatility of the simulations has resulted in a remarkable scientific impact: a conservative estimate based on searches using the NASA Astrophysics Data System is that data and/or data products from these models feature in over one thousand astrophysics research articles published since 2014. 

It is imperative to remember that much of galaxy formation and evolution modelling remains distantly removed from fundamental, ab initio theory. There is no guarantee that the reproduction of a particular observable represents a unique solution, i.e.\ different subgrid models may yield similarly-successful outcomes. It is possible that the approximations necessitated by subgrid models may deliver success only at a particular resolution, or only at the expense of a failure elsewhere. Simulations are therefore often better suited to offering qualitative insight rather than quantitative predictions: some of the most instructive outcomes from simulation suites have stemmed from varying the efficiency of physical mechanisms to isolate their influence (and hence indicate \textit{why} a simulation reproduces a particular observable), or from a failure to reproduce particular observational measurements and so illuminate a fundamental shortcoming of the implemented physics, or even the adopted cosmogony. 

\begin{marginnote}[]
\entry{$\Lambda$CDM}{$\Lambda$-Cold Dark Matter}
\end{marginnote}

The chief objective of this review is therefore to offer a critical assessment of the improved understanding of the formation and evolution of the galaxy population fostered by the current generation of state-of-the-art hydrodynamical (or magneto-hydrodynamical) simulations of the $\Lambda$-cold dar matter ($\Lambda$CDM) cosmogony. To this end, we highlight the key successes of the simulations that we believe will endure, and the shortcomings or absences of consensus that present outstanding challenges to be addressed by future models. Where possible, we offer candid explanations for the origin of shortcomings, which we hope will be particularly helpful to non-specialists and new practitioners. We focus on simulations that follow reasonably representative cosmological volumes ($L \simeq 100$ comoving $\Mpc$, hereafter $\cMpc$) at fixed resolution, and thus yield simulated \textit{populations} of galaxies. These simulations are complementary to zoom simulations of individual galaxies, but we focus on the former because in general it is simpler to compare their outcomes with observations and to characterise their convergence behaviour (Section \ref{sec:convergence}), and because one cannot confuse trends due to resolution with those due to galaxy or halo mass, which is a danger when analysing suites of zoom simulations that maximise the resolution at each mass scale. A major advantage of this type of simulation is the diversity of the lines of enquiry that they enable, so we necessarily restrict ourselves to reviewing their more fundamental outcomes.

% Definitions
\begin{marginnote}[]
\entry{SPH}{Smoothed Particle Hydrodynamics}
\entry{AMR}{Adaptive Mesh Refinement}
\end{marginnote}
We explore results primarily (but not exclusively) from the \texttt{EAGLE}, \texttt{Horizon-AGN} \citep{Dubois2012, Kaviraj2017}, \texttt{IllustrisTNG} and \texttt{SIMBA} projects, which are now somewhat mature simulation suites that have been studied in detail (key numerical details of these simulations are given in Table~\ref{tab1}). The flagship simulation of each project follows the evolution of a cosmological volume of $L\sim 100\cMpc$ with baryonic mass (spatial) resolution of $\sim 10^{6-7}\,\Msun$ ($\sim 1$ proper $\kpc$, hereafter $\pkpc$ in the ISM), and yields a galaxy population that is, to a greater or lesser degree, broadly realistic. They span the range of hydrodynamics solvers widely in use [Modern smoothed particle hydrodynamics (SPH), adaptive mesh refinement (AMR), and hybrid Lagrangian-Eulerian approaches, see Section \ref{sec:methods}], and adopt a diverse range of subgrid implementations of baryonic physics, each differing from the others in at least several major aspects. Of these, only \texttt{IllustrisTNG} solves the equations of magneto-hydrodynamics, the others use pure hydrodynamics. \texttt{EAGLE} and \texttt{IllustrisTNG} are the most well-studied and readily-accessible of these simulations, and consequently feature more frequently in this review. Our focus on these simulation suites is to provide illustrative examples, and should not be misinterpreted as implicit criticism of other simulation campaigns, nor dismissal of their successes. 

The review is structured as follows. We discuss the methods used in the execution, calibration and analysis of galaxy population simulations in Section \ref{sec:methods}. We review in Sections \ref{sec:key_properties} and \ref{sec:scaling_relns} the key properties of simulated galaxy populations, and scaling relations connecting diverse properties of simulated galaxies to their stellar mass, respectively. Section \ref{sec:cosmic_gas} focuses on the properties of the gaseous environments of simulated galaxies, and Section \ref{sec:environment} examines environmental influences. We discuss likely future directions for the discipline in Section \ref{sec:future}, and provide a brief summary of our conclusions in Section \ref{sec:summary}.

\begin{table}[h]
\tabcolsep7.5pt
\caption{Details of simulations used for illustrative examples}
\label{tab1}
\begin{center}
\begin{tabular}{@{}l|l|l|l|l|l@{}}
\hline
Simulation & Hydro method & Boxsize  & DM particles  & Baryonic mass        & Baryonic spatial \\
           & & [cMpc]   & per dimension & resolution [$\Msun]$ & resolution$^{\rm a}$ [pkpc] \\
\hline
\texttt{EAGLE}$^{\rm b}$        & Modern SPH  & 100.0  & 1504 & $1.9\times 10^6$ & 0.70\\
\texttt{Horizon-AGN}            & AMR         & 142.0  & 1024 & $1.0\times 10^7$ & 1.00 \\
\texttt{IllustrisTNG}$^{\rm c}$ & Moving mesh & 110.7  & 1820 & $1.4\times 10^6$ & 0.19 \\
\texttt{SIMBA}$^{\rm d}$        & MFM         & 147.0  & 1024 & $1.8\times 10^7$ & 0.74 \\
\hline
\end{tabular}
\end{center}
\begin{tabnote}
$^{\rm a}$Maximum size of the gravitational softening for gas.\\ $^{\rm b}$Except where explicitly stated, we consider the Ref-L100N1504 simulation from the \texttt{EAGLE} suite.\\ $^{\rm c}$Except where explicitly stated, we consider the TNG-100 simulation from the \texttt{IllustrisTNG} suite.\\ $^{\rm d}$Except where explicitly stated, we consider the m100n1024 simulation from the \texttt{SIMBA} suite
\end{tabnote}
\end{table}

\section{Methods}
\label{sec:methods}

\subsection{Initial \& boundary conditions} 
\label{sec:IC}

Cosmological simulations begin from initial conditions that specify the fluctuations in the density field at an epoch at which there is no significant non-linearity, generally $z \gtrsim 100$. From this initial state, the equations of motion can be integrated forwards into the non-linear regime numerically. The initial matter power spectrum (which specifies the density contrast relative to the mean density as a function of spatial scale) is usually expressed as the product of an initial spectrum (with random phases) resulting from inflation, and a transfer function (which can be calculated with a Boltzmann solver such as CAMB; \citealt{Lewis2011}) that represents the subsequent linear evolution of each mode. Cosmological simulators are therefore in the fortunate position of having initial conditions that are well constrained by observations of the cosmic microwave background (CMB) radiation at $z\approx1100$. 

Generating Gaussian random fields with a specified power spectrum, a process first detailed by \citet{Efstathiou1985}, has evolved into a specialist discipline. We only briefly summarise the process here and encourage those with a particular interest to read the work of, for example, \citet{Jenkins2010,Jenkins2013,Hahn2011}, and \citet{Hahn2021}. The process comprises two key stages to realize the appropriate density fluctuations: the creation of uniform particle distribution throughout the simulation volume, followed by the application of displacements to positions and velocities of the particles or, alternatively, small adjustments to the masses of particles. The unperturbed particle distribution is usually constructed by tiling a cubic grid or a `glass' distribution \citep[][]{White1994}. Random particle distributions are unsuitable because they exhibit a white noise power spectrum that, even in the absence of the intended displacements, fosters rapidly-growing non-linear structure. The displacements are calculated with linear or low order perturbation theory. Periodic boundary conditions are applied to opposing faces of the volume, ensuring that its mean density remains fixed and that no artificial boundaries are imposed.

The finite size of the simulation domain imposes an upper limit to the spatial scale of the fluctuations that can be sampled within it (correspondingly, there is a minimum wavenumber of $k=2\pi/L$). As large scale fluctuations seed the formation of rare features in the cosmic large-scale structure, such as galaxy clusters and large voids, the emergent space density of such features is underestimated \citep{Bagla2006,Reed2007}. Volumes of $L\simeq 100\cMpc$ are too small to realise galaxy clusters (whose present-day space density is $< 10^{-6}\perMpc$), such that detailed examination with (magneto-)hydrodynamical simulations of the galaxy population \textit{within} them requires the use of zoomed initial conditions \citep[e.g.][]{Bahe2017,Cui2018,Tremmel2019}. The scarcity of rare features in finite simulation volumes also makes them sensitive to cosmic variance, as different realisations of the same initial density fluctuations can result in statistically significant differences in the emergent large-scale structure.

\subsection{Gravitational \& hydrodynamical evolution} 
\label{sec:grav_hydro}

Modelling the formation of the galaxy population entails solving the partial differential equations that govern the temporal evolution of the cosmic matter and radiation fields. Simulations of the galaxy population apply numerical techniques to solve the equations governing the gravitational evolution of matter, the hydrodynamical evolution of gas and, in some cases, the interaction of gas with evolving radiation and magnetic fields. Other physical processes are treated with subgrid models. Since the key numerical techniques are documented, and their advantages and shortcomings discussed, in detail elsewhere \citep[see e.g.][]{Springel2010_SPHreview, Price2012, Teyssier2015}, we only briefly discuss them. We focus primarily on a discussion of subgrid methods (Section \ref{sec:subgrid_methods}), because these dominate systematic uncertainties at the resolution of the simulations examined here.

Dark matter is most commonly treated as a collisionless fluid \citep[but for self-interacting treatments, see e.g.][]{Dave2001,Robertson2017}, whose evolution in the continuum limit is described by the collisionless Boltzmann equation (CBE), under the influence of the gravitational potential given by Poisson's equation. The potential is assumed to be Newtonian, because velocities on resolved scales are non-relativistic. The high dimensionality of the CBE necessitates solving the coupled equations using a finite set of $N$-body tracer particles that sample the fluid's phase space distribution. The $\mathcal{O}(N^2)$ scaling of the computational cost of solving Poisson's equation (stemming from the long-range nature of gravity that requires consideration of $N-1$ contributions to the potential at each particle), can be reduced to a scaling on the order of $\mathcal{O}(N\log N)$ by approximating the contribution from distant particles. This is achieved via multipole expansion \citep[e.g.][]{Barnes1986,Carrier1988}, and/or by mapping the tracer distribution to a mesh and solving in Fourier space using fast transform methods \citep[e.g.][]{Hockney1981}. To prevent the unphysical scattering of close particle pairs, the gravitational force is softened on small scales using a kernel function, for which forms of varying complexity have been proposed \citep[e.g.][]{Monaghan1985,Wendland1995}. It is common to adopt a softening scale that is fixed in comoving units (e.g.\ $\simeq 1/25$ of the mean interparticle separation) and limited to a maximum proper size to ensure that the internal structure of dark matter haloes can be resolved at late cosmic epochs \citep{Power2003}. 

Cosmic gas is assumed to be ideal, collisional, inviscid, and non-conducting, enabling its dynamics to be described by the Euler equations rather than the more general Navier-Stokes equations. Traditionally, solution of the equations has been via two distinct approaches for discretising the fluid: either in volume (the Eulerian approach used by mesh-based schemes) or in mass (the Lagrangian approach used by particle-based schemes). Each approach has well-advertised shortcomings that contribute to inconsistent results when applied to relatively simple cosmological structure formation problems involving non-radiative hydrodynamics \citep[e.g.][]{Frenk1999, Sembolini2016}.  Eulerian methods are the de facto standard for many computational fluid dynamics problems, but the dynamic range needed to resolve galaxies in cosmological volumes demands the added complexity of adaptively-refined meshes \citep[AMR, e.g.][]{Abel2002}, which may still fail to adequately capture the gravitational collapse of low-contrast fluctuations at early times \citep{OShea2005} and cause over-mixing of gas with differing entropies, e.g.\ in the cores of galaxy clusters \citep[][]{Wadsley2008,Mitchell2009}. The inherent Galilean non-invariance of mesh methods \citep[][]{Tasker2008} is particularly undesirable because the relative velocities of galaxies are generally much greater than the sound speed of the gas bound to them. 

SPH \citep[][]{Lucy1977,Gingold1977} samples the fluid with tracer particles, naturally adapting the resolution within overdense structures. SPH is hence well suited to following the hierarchical growth of structure, and enables the self-gravity of the gas to be treated in an identical fashion to the dark matter, but requires the inclusion of an artificial viscosity term to capture shocks \citep[][]{Monaghan1997}, which are resolved poorly. Traditional SPH implementations suffer from multivalued particle pressure at contact discontinuities that create unphysical surface tension and inhibit phase mixing \citep[e.g.][]{Agertz2007}, though this problem can be mitigated via the use of various adaptations \citep[][]{Ritchie2001,Price2008,Hopkins2013,Wadsley2017,Borrow2022a}, which have become collectively known colloquially as `modern' or `corrected' SPH.

In the last decade, several software packages have emerged that treat cosmic gas with sophisticated schemes that seek to capture the advantages of both Lagrangian and Eulerian approaches, i.e. continuous adaptability of resolution and geometry of the (magneto-)hydrodynamics calculation, Galilean invariance, accurate treatment of fluid mixing, and high-fidelity shock capturing without the use of artificial viscosity. These schemes (sometimes referred to as `arbitrary Lagrangian-Eulerian', or ALE) typically use a mesh that deforms and moves with the fluid flow \citep[`moving mesh', e.g.][]{Springel2010_arepo,Vandenbroucke2016}, or solve the Riemann problem without a mesh \citep[e.g. meshless finite volume, MFV, or meshless finite mass, MFM;][]{Hopkins2015} (though note that MFM is not strictly an ALE method because it uses resolution elements with fixed mass). These approaches have proven successful for particular problems, but simultaneously realising the benefits of Eulerian and Lagrangian methods generally incurs a marked increase in computational cost and memory footprint. 

Non-standard refinement criteria have been used with both AMR and moving mesh schemes, basing control of the resolution of the hydrodynamics calculation on conditions other than the fluid density. This has enabled examination of fluid flows in greater detail, for example, around supermassive black holes (SMBHs, \citealt{Curtis2015}) or in the gaseous haloes around galaxies (the circumgalactic medium (CGM), \citealt{Hummels2019,Peeples2019,vandeVoort2019}).
 
\begin{marginnote}
\entry{SMBH}{Supermassive black hole}
\entry{CGM}{Circumgalactic medium}
\end{marginnote}

\subsection{Subgrid methods}
\label{sec:subgrid_methods}

\begin{figure}[h]
\includegraphics[width=5in]{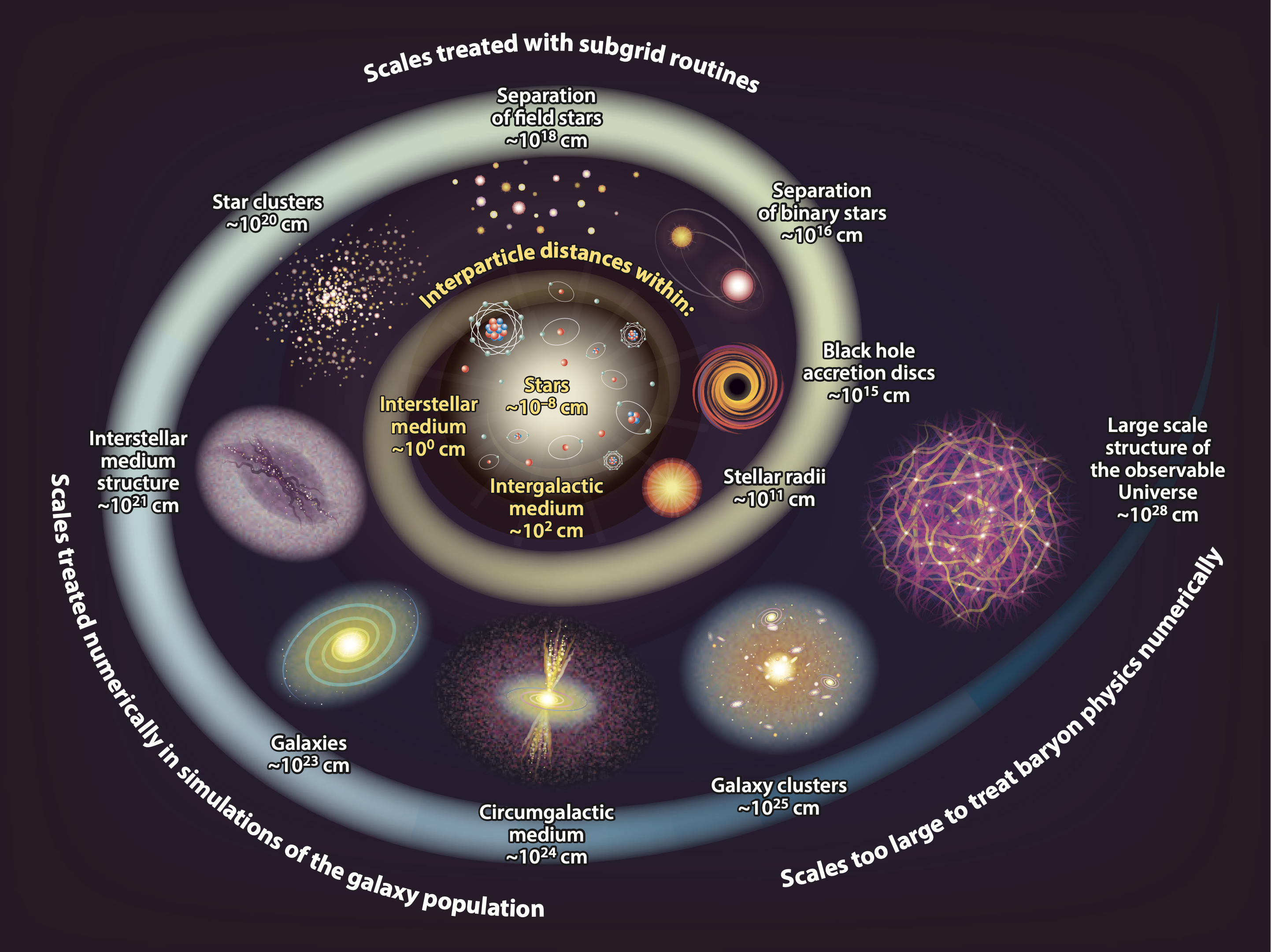}
\caption{Schematic illustration of the extreme dynamic range of the length scales of the physical processes influencing the formation and evolution of galaxies. Cosmological, hydrodynamical simulations are able to follow cosmic structure formation up to $\mathcal{O}$(100~Mpc) scales and have to resort to subgrid models on small scales. Future simulations of the galaxy population will push both of these boundaries, increasing their dynamic range, but the need for subgrid models will remain for years to come.}
\label{fig:scales}
\end{figure}

The need to uniformly sample representative cosmic volumes restricts the feasible resolution of galaxy population simulations. Let us consider a brief example: assuming $\Omega_{\rm b} = 0.05$ and $H_0=70$~km~s$^{-1}$Mpc$^{-1}$, uniformly sampling the gas fluid in a cubic volume of $L=100\ \cMpc$ with tracers of mass $10^5\,\Msun$ (comparable to the stellar mass of ultra-faint dwarf galaxies) requires $\simeq 68$ billion fluid elements ($N=4082^3$). Simulations including a broad suite of baryonic physics can require $\sim 1$kB per fluid element, incurring a total memory footprint for the baryonic part of the calculation of $\simeq 68$TB. Assuming an equal number of $N$-body particles for the dark matter, at 100 bytes per particle, brings the footprint to $\simeq 75$TB. With current high-performance computing facilities typically having 2-4GB of memory per core, our example would require execution on between $20,000$ and $40,000$ cores. Load balancing relatively high-resolution simulations at this scale remains extremely challenging, despite the development of sophisticated schemes for this purpose \citep[see e.g.][]{Menon2015,Schaller2016,Weinberger2020}, largely because of the extreme dynamic range of the timestep hierarchy. As a result, calculations of this scale need to occupy these large core counts for prohibitively long periods (often several months). The challenge of accommodating the germane physical processes into simulations is illustrated schematically by Figure~\ref{fig:scales}, which shows their characteristic length scales, and highlights that subgrid models remain a critical component of simulations of the galaxy population for the foreseeable future. We review the key physical processes treated with subgrid methods below.

\subsubsection{Radiative processes in cosmic gas}
\label{sec:subgrid_radiative}

Radiative processes enable cosmic gas to dissipate its internal energy. In the absence of an incident radiation field, the ionisation balance and cooling rate of diffuse cosmic gas is governed by two-body processes (e.g. collisional excitation and ionization, collisional recombination, and free–free emission). The cooling rate due to these processes scales as the square of the gas density, and the contribution of each elemental species scales linearly with its abundance, enabling convenient tabulation of colliosional ionisation equilibrium (CIE) cooling rates as a function of temperature and composition \citep[e.g.][]{Dalgarno1972}. CIE rates (plus inverse Compton cooling due to the CMB) for gas of primordial composition were adopted by the first generation of hydrodynamical simulations of galaxies \citep[e.g.][]{Katz1996}, and remain in use in a limited number of modern simulations [e.g.\ \texttt{MassiveBlack-II} \citep{Khandai2015}, \texttt{Romulus} \citep{Tremmel2015}]. CIE rates for metal enriched gases can differ from primordial rates by an order of magnitude \citep[e.g.][]{Boehringer1989,Sutherland1993} and are more widely used. However, it should be noted that uncertainties in nucleosynthetic yields (see Section \ref{sec:subgrid_stellar_evolution}) propagate into uncertainties in gas cooling rates. Metal cooling allows enriched ionised gas to cool below $10^4\,\K$, though in many simulations such cooling is effectively disabled (at least for dense gas, $n_{\rm H} \gtrsim 0.1~\percc$) by the use of an effective equation of state or pressure floor (see Section \ref{sec:subgrid_ism}). CIE rates for enriched gases (with abundance ratios similar to those of the solar neighborhood) are used by, for example, the \texttt{Horizon-AGN} and \texttt{BlueTides} \citep{Feng2016} simulations.
\begin{marginnote}
\entry{CIE}{Collisional ionisation equilibrium}
\end{marginnote}

Cosmic radiation fields can `overionise' gas relative to CIE, reducing the cooling rate or even yielding radiative heating \citep[e.g.][]{Efstathiou1992,Gnedin2012}. The most dramatic example is the epoch of reionisation (EoR), when intergalactic neutral hydrogen (\HI) is (re-)ionised by radiation from the first galaxies and active galactic nuclei (AGN). The accurate treatment of this process in simulations, which is necessary to interpret a diverse range of observable properties of the first generations of galaxies and intervening IGM, requires the inclusion of explicit, and computationally-expensive, radiative transfer calculations. Simulations of the galaxy population instead generally adopt a temporally-evolving metagalactic UV/X-ray background (UVB) radiation field \citep[for popular parametrisations see][]{Haardt1996,Haardt2001,Haardt2012,FaucherGiguere2009}, switched on at the `instantaneous' reionisation redshift inferred from the Thomson scattering optical depth of the CMB. The field is assumed to be spatially uniform, because the clustering scale of the photon sources is shorter than the mean free path of the photons in the diffuse IGM \citep[e.g.][]{Zuo1992}, though this approximation is less applicable close to galaxies. Cooling rates are then computed assuming the gas to be optically thin and in ionisation equilibrium, using spectral synthesis codes such as \textsc{cloudy} \citep[][]{Ferland2017}. 
\begin{marginnote}
\entry{AGN}{Active galactic nuclei}
\end{marginnote}

The photoionisation rate scales linearly with density, so radiative cooling rates considering photoionisation and CIE must be tabulated as a function of density, temperature, redshift and composition \citep[e.g.][]{Smith2008}. The contribution of each species is additive, enabling `element-by-element' tabulation and affording the flexibility to model cooling in gases with abundance ratios that deviate from those of the solar neighborhood. \citet{Wiersma2009a} provide such a tabulation, for the set of 11 most-important radiative coolants (H, He, C, N, O, Ne, Mg, Si, S, Ca and Fe) exposed to the \citet{Haardt2001} UVB model. These tables are used by the \texttt{EAGLE}, \texttt{Illustris}/\texttt{IllustrisTNG}, \texttt{FABLE}  and \texttt{Magneticum}  simulation suites, though only \texttt{EAGLE} and \texttt{Magneticum}  exploit the element-by-element nature of the tables; the other suites scale the cooling rate as a function of metallicity, assuming solar neighborhood abundance ratios.

\begin{marginnote}
\entry{UVB}{UV/X-ray background}
\entry{ISM} {Interstellar medium}
\end{marginnote}
As the column density of cosmic gas approaches a vlaue characteristic of Lyman-limit systems ($N_{\rm HI} \gtrsim 10^{17}\percmsq$), it becomes optically thick, attenuating the incident radiation flux and increasing the cooling rate relative to the optically thin case. The corresponding volume density of gas for which self-shielding is important is $n_{\rm H} \gtrsim 10^{-2}\,\percc$ \citep[e.g.][]{Furlanetto2005}, characteristic of the interstellar medium (ISM). The reduction of the UVB photoionisation rate in self-shielded gas can be approximated using fitting functions calibrated against full radiation transport calculations \citep[e.g.][]{Gnedin2009,Rahmati2013}. This approximation is applied on-the-fly (for $z<6$) by the \texttt{Illustris} and \texttt{IllustrisTNG} simulations. The self-shielding approximation of \citet{Rahmati2013} is also used by the \textsc{Grackle} library \citep[last described by][]{Smith2017}, which computes cooling rates for interstellar gas using a primordial chemistry reaction network, in tandem with tabulated rates for metal species in the optically-thin regime. The primordial network also accounts for deviations from ionisation equilibrium. This library is used by the \texttt{SIMBA} simulations. 

Local radiation fields in the ISM can dominate over the UVB, which further complicates the calculation of cooling rates \citep[e.g.][]{Schirber2004,MiraldaEscude2005}. Molecules and dust grains within the ISM attenuate the incident radiation spectrum (in dissimilar ways), and also contribute to the cooling rate \citep[e.g.][]{Omukai2005}. \citet{Ploeckinger2020} provide cooling tables for gas, shielded by itself and dust grains, in ionisation equilibrium with radiation from both a spatially-uniform, temporally-evolving UVB and an interstellar radiation field of variable intensity, but with spectral shape fixed to that of the Milky Way \citep{Black1987}.

\subsubsection{Element abundance evolution}
\label{sec:subgrid_stellar_evolution}

Besides influencing radiative cooling rates, element abundance patterns in cosmic gas and stars encode a wealth of astrophysical information. Elements heavier than lithium (`metals') were produced by nucleosynthetic processes within stars and supernovae (SNe) (and in some cases, kilonovae), with some fraction of the synthesised elements returned to the local ISM as ejecta. The enriched gas can then be incorporated into future generations of stars or transported into the CGM or IGM by galaxy-scale outflows. When galaxies fall into more massive haloes, such as groups and clusters of galaxies, their enriched ISM can be stripped by the ram pressure it experiences from the CGM or intracluster medium (ICM). 

\begin{marginnote}
\entry{SN(e)}{Supernova(e)}
\entry{ICM}{Intracluster medium}
\entry{SSP}{Simple stellar population}
\entry{SNeII}{Type II supernovae}
\entry{SNeIa}{Type Ia supernovae}
\entry{AGB}{Asymptotic giant branch}
\end{marginnote}

Modern schemes for distributing and transporting the elements synthesised by stellar populations (`chemodynamics') remain similar to the first generation of implementations (\citealp[e.g.][]{Theis1992,Steinmetz1994,Carraro1998}. Stellar particles, each assumed to represent a mono-age simple stellar population (SSP), donate enriched mass to neighboring fluid tracers. Elements then share the destiny of the tracer to which they are donated, enabling their transport and dispersal by gas flows, and incorporation into subsequent generations of stars, in a fashion consistent with the adopted hydrodynamics scheme. Early schemes focused on enrichment by Type II SNe (SNeII), whose progenitors have short lifetimes ($\lesssim 30\Myr$, comparable to the sound-crossing time of fluid elements representing the diffuse ISM) and whose nucleosynthetic products can be distributed immediately upon the formation of the SSP forms (the `instantaneous recycling approximation'). Although the total metallicity is generally dominated by elements synthesised by SNeII, modelling the abundance of elements such as iron, carbon and nitrogen requires consideration of element release by Type Ia SNe (SNeIa) and long-lived stars that experience an asymptotic giant branch (AGB) phase. Modern simulations therefore generally adopt subgrid models that follow the timed release of individual elements from multiple nucleosynthetic channels, e.g.\ \citet{Oppenheimer2008}, used by \texttt{SIMBA}; \citet{Tornatore2007} used by \texttt{Magneticum} ; and \citet{Wiersma2009b} used by \texttt{EAGLE} and \texttt{Illustris}/\texttt{IllustrisTNG}. 

The progenitors of AGB stars and SNeII are, respectively, intermediate ($0.8 \lesssim M \lesssim 8\,\Msun$) and high ($M \gtrsim 8\,\Msun$) mass stars. An initial mass function (IMF) must therefore be specified in order to set the relative number of these progenitors (it is also required to set the energetics of stellar feedback, see Section\,\ref{sec:subgrid_feedback}). It is typically assumed to follow the form deduced from solar neighborhood number counts \citep[e.g.][]{Chabrier2003}, with the lower mass cut-off being the hydrogen mass burning limit ($\simeq 0.07-0.09\,\Msun$) and the upper mass cut-off reflecting the (more uncertain) maximum observed mass of stars. Typically limits of $0.1\,\Msun$ and $100\,\Msun$ are adopted. Uncertainties in the form and limits of the IMF translate directly into uncertainty on the cosmic metal budget. Mass loss during the AGB phase occurs in a brief phase at the end of the main sequence lifetime. Lifetimes are challenging to constrain observationally and are generally inferred from stellar evolution models \citep[e.g.][]{Romano2005}. The models concur that lifetimes are a strongly-decreasing function of stellar mass, with some inferring a weak dependence on metallicity. The intermediate mass progenitors of AGB stars have lifetimes of $10^8 - 10^{10}\,\yr$, which are long compared to the sound crossing time of ISM fluid elements (and hence the timesteps on which dense gas is advanced), but are in general shorter than, or comparable to, the dynamical timescales of galaxies. The mass dependence of stellar lifetimes therefore has a tangible influence on element release. 

SNeIa are thermonuclear explosions resulting from binary star evolution. Their progenitors remain ill-constrained, with the most plausible cases being the accretion of mass onto a Chandrasekhar mass white dwarf from a non-degenerate companion, or the merger of two white dwarfs \citep[e.g.][]{Maoz2014}. The SNIa rate resulting from either scenario is thus complicated by the ill-constrained properties of binaries, e.g.\ their mass fraction, mass function, and initial separation, however the involvement of at least one white dwarf dictates that the binary progenitor is typically old ($\gtrsim 10^9\,\yr$). Forward modelling the SNIa rate as a function of an SSP's age is therefore subject to many uncertainties \citep[e.g.][]{Greggio1983,Greggio2005}, fostering the alternative approach of assuming an empirical delay function for the SSP's SNIa rate, which integrates to unity at $t=\infty$ and whose free parameters can be calibrated against cosmic SNIa observations \citep[see e.g.][]{Mannucci2006,Wiersma2009b}. 

The nucleosynthetic yields of SNeII, SNeIa, and AGB stars represent a significant systematic uncertainty for chemodynamical modelling. The uncertainties stem from the complexity of the physics involved, such as rotation in SNII progenitors, or the efficiency of convective envelope burning in AGB stars, the post-explosion time at which the yields are quoted (synthesised isotopes can have radioactive decay timescales shorter than the expansion timescale of the ejecta), or inconsistencies in the assumed mass ranges of progenitors. Nucleosynthetic calculations remain challenging to reconcile with the observed abundances of stars in the Galaxy, often motivating ad-hoc and element-specific rescaling factors (at the factor $\simeq 2$ level) to the resulting theoretical yields \citep[e.g.][]{Francois2004,Portinari2004}. As such, the absolute element abundances predicted by simulations must be considered uncertain by at least a similar factor. 

Mixing and diffusion are further significant sources of uncertainty for chemodynamical evolution. Some degree of overmixing is generally expected in mesh-based simulations, because fluids are implicitly mixed on scales smaller than that of the smallest cells, but excellent numerical convergence behaviour can be demonstrated with moving mesh treatments of hydrodynamics \citep{vandeVoort2020}. Conversely, SPH (particularly traditional schemes) tends to underestimate mixing because of its inability to resolve dynamical instabilities \citep[e.g.][]{Agertz2007}. In the absence of an explicit diffusion treatment, metal mixing in SPH simulations is further underestimated because metals are `stuck' to particles, resulting in poor sampling of fluids with a small but non-zero metallicity. The problem can be mitigated by the inclusion of a subgrid diffusion treatment \citep[e.g.][]{Wadsley2008,Greif2009}, but the appropriate coefficients are unclear \citep[e.g.][]{Garnier2009}. \citet{Wiersma2009b} argue that a simpler strategy to mitigate the metal sampling problem in SPH simulations is to use kernel-smoothed abundances. Irrespective of the hydro solver, undermixing can be exacerbated if, as is commonplace, sources of small-scale turbulence are neglected.

\subsubsection{Subgrid models of the ISM and star formation} 
\label{sec:subgrid_ism}

Perturbations in a self-gravitating medium can grow only if their wavelength exceeds the \citet{Jeans1928} length. Accurate numerical modelling of gravitationally collapsing systems therefore requires that the Jeans scales are adequately resolved: in hydrodynamical settings, this translates into the need to resolve pressure forces on the scale at which self-gravity begins to dominate. In Lagrangian simulations the condition is most naturally expressed in terms of the ratio of the particle mass and the Jeans mass \citep{Bate1997}, and in Eulerian counterparts in terms of the ratio of the cell size and the Jeans length \citep{Truelove1997}.

The thermogravitational collapse of the warm, diffuse phase of the ISM ($T\sim 10^4\,\K$, $n_{\rm H}\sim 10^{-1}\,\percc$) to the cold, dense phase ($T \lesssim 10^2\,\K$, $n_{\rm } \gtrsim 10^2\,\percc$) occurs quickly \citep[][]{McKee1977}. Simulations that seek to resolve the molecular ISM are therefore faced with demanding resolution requirements. The Jeans scales depend on both the density and temperature of the gas ($M_{\rm J} \propto n_{\rm H}^{-1/2}T^{3/2}$, $L_{\rm J} \propto n_{\rm H}^{-1/2}T^{1/2}$), so an increase of (at least) three decades in density and a decrease of (at least) two decades in temperature corresponds to reductions in the Jeans mass from $M_{\rm J}\sim 10^7\,\Msun$ to $M_{\rm J}\sim 10^2\,\Msun$, and the Jeans length from $L_{\rm J}\sim 1\,\kpc$ to  $L_{\rm J}\sim 1\pc$. As noted earlier, appealing to such high resolution in representative cosmological volumes requires the use of very short timesteps and incurs a very large memory footprint, necessitating execution with large core counts that cannot be effectively exploited by the current generation of simulation codes. The pressure structure of the multiphase ISM therefore cannot be accurately resolved, typically requiring that a subgrid model is either used to predict the pressure structure as a function of the gas density \citep[e.g.][]{Yepes1997,Springel2003}, or to impose the structure `by hand' as a pressure floor \citep[e.g.][]{Machacek2001,Robertson2008} or as an equation of state \citep[e.g.][]{Schaye2008}. 

\begin{marginnote}
\entry{SFR}{Star formation rate}
\end{marginnote}
In the absence of a numerical treatment of the cold ISM, star formation must be implemented with models whose efficiency is calibrated to reproduce observed scaling relations averaged over suitably large spatial scales ($\sim 1\kpc$). The most commonly adopted form is that of a \citet{Schmidt1959} law, for which the star formation rate (SFR) density ($\dot{\rho}_\star$) scales linearly with gas density ($\rho_{\rm g}$) over a dynamical time ($t_{\rm dyn}$): $\dot{\rho}_\star \propto \epsilon_\star \rho_{\rm g}/t_{\rm dyn}$. Here, $\epsilon_\star$ is the efficiency per dynamical time which is calibrated so that, operating in tandem with the density threshold above which star formation is allowed to proceed, the simulation reproduces the observed Kennicutt-Schmidt (KS) star formation law \citep{Kennicutt1998}. Since the latter is based on surface densities rather than volume densities, the calibration depends on the assumed equation of state. \citet{Schaye2008} therefore introduced an alternative scheme that, under the assumption of vertical hydrostatic equilibrium in the ISM, enables the KS law to be expressed as a pressure law, eliminating the dependence of the star formation law on the equation of state and replacing the calibrated parameters with observables. Alternatively, \citet{Feldmann2023} presented results from the \texttt{FIREbox} simulation of a (small, $L = 22.1\cMpc$) cosmological volume at $z=0$, whose star formation model \citep{Hopkins2018} does not impose an efficiency per free-fall time, instead aiming to recover the observed effective efficiency via the influence of the model's implemented ISM and feedback physics.

\subsubsection{SMBH seeding and growth}
\label{sec:subgrid_bh}

SMBHs are represented by collisionless sink particles, subject only to gravity. They are usually seeded at the centres of haloes (which do not already have a seed) identified on-the-fly by periodically applying a group-finder algorithm to the dark matter particle distribution \citep[e.g.][]{DiMatteo2008}, or in high density gas whose properties satisfy a number of criteria \citep[e.g.][]{Dubois2012}. There is no concensus concerning the formation mechanism of SMBHs \citep[see e.g.][]{Volonteri2021}, but popular theories posit that their formation mass is below the resolution limit of simulations of the galaxy population, requiring that sink particles carry a subgrid SMBH mass (used by subgrid routines concerning SMBH evolution) in addition to their dynamical mass. Once the former exceeds the latter, the subgrid and dynamical masses grow in concert. Sink particles with masses less than or comparable to the mass of neighboring resolution elements do not experience a realistic dynamical friction force, so it is common to `pin' them to the halo's centre of potential, or migrate them towards it. This practice has been shown to have significant consequences for the evolution of simulated galaxies \citep{Bahe2022}.

SMBHs grow via gas accretion and mergers with other SMBHs. The ambient gas accretion rate, $\dot{m}_{\rm accr}$ is generally assumed to be at the minimum of the Eddington and Bondi-Hoyle-Lyttleton rates \citep[e.g.][]{Bondi1944}, or a modified version of the latter, though recent models also explicitly consider the influence of the angular momentum of the gas \citep[e.g.][]{AnglesAlcazar2017}. The Bondi rate is estimated from the sink particle's ambient gas properties, necessarily on spatial scales that are much larger than accretion discs, such that the simulations treat the physics of accretion onto SMBHs in a very simplistic fashion \citep[see e.g.][]{Shlosman1990}. The SMBH grows at a rate $\dot{m}_{\mathrm{BH}}=\dot{m}_{\rm accr}(1-\epsilon_{\rm r})$, where $\epsilon_{\rm r}$ is the SMBH's radiative efficiency, generally assumed to follow the mean value for radiatively-efficient accretion onto a Schwarzchild BH, $\epsilon_{\rm r}=0.1$ \citep{Shakura1973}. Older simulations based on the popular model of \citet{Springel2005_quasars} boosted the Bondi rate (usually by a factor of $\alpha=100$) to compensate the underestimate of the gas density near the Bondi radius, though modern simulations tend to apply either a density-dependent correction as advocated by \citet{Booth2009}, or no correction because it is generally only important early in the SMBH's growth history, making it degenerate with the choice of seed mass. A more sophisticated approach possible in mesh-based simulations is to refine the resolution of the simulation in the vicinity of SMBHs to ensure the Bondi radius is resolved \citep{Curtis2015}. 

\subsubsection{Feedback processes}
\label{sec:subgrid_feedback}

The ratio of galaxies' stellar mass to halo mass, as inferred for example by sub-halo abundance matching \citep[e.g.][]{Conroy2006}, exhibits a characteristic peak at a halo mass of $\sim10^{12}\,\Msun$. The efficiency with which haloes convert their cosmic `share' of gas into stars decreases towards lesser and greater halo mass scales. This is widely interpreted as a signal that galaxy growth is primarily regulated (or even quenched) by separate feedback mechanisms in these two regimes: the formation and evolution of stellar populations in low-mass galaxies, and accretion onto SMBHs in massive galaxies. The injection of energy by these mechanisms occurs on numerically unresolved scales, so simulations must approximate their macroscopic effects with subgrid models. 

Stellar populations inject energy and momentum into the ISM via stellar winds, radiation and SNe, and thus potentially disrupt star-forming molecular clouds, create turbulence, and drive gas out of galaxy discs. The simplest subgrid treatment sums the energy liberated at each timestep by a stellar particle (representing an SSP), and injects it thermally by raising the internal energy of the particle's neighboring fluid elements. However, because these elements are orders of magnitude more massive than SNe ejecta, the injected energy (canonically $E_{\rm SN} \sim 10^{51}\,{\rm erg/SN}$) is only sufficient to heat individual fluid elements to $\sim 10^{5}\,\K$. In this regime the gas cooling time is short compared to its sound crossing time, leading to catastrophic artificial losses that preclude the formation of an adiabatic, energy-conserving blast wave \citep[e.g.][]{Katz1996,DallaVecchia2008,Creasey2011}. 

Popular solutions are to temporarily disable the cooling of gas, either by hand \citep[e.g.][]{Gerritsen1997,Stinson2006} or using an additional subgrid model to account for unresolved ISM physics \citep{Keller2014}; to heat gas particles stochastically by temperature increments $\Delta T \gg 10^5\,\K$ \citep{Kay2003,DallaVecchia2012}; or to inject (some of) the energy in kinetic form \citep[e.g.][]{Navarro1993,Springel2003,Dubois2008}. Each approach has pros and cons but, for coupling efficiencies of order unity, they all in principle enable (low-mass) galaxies to drive outflows and achieve self-regulated growth. The kinetic approach affords the freedom to specify explicitly both the initial velocity and mass loading ($\eta = \dot{M}_{\rm outflow}/\dot{M}_\star$) of the wind, enabling calibration of the wind model against the outcomes of higher resolution zoom simulations with more sophisticated ISM models, an approach adopted by \texttt{SIMBA}. The popular \citet{Springel2003} implementation, which has been used in adapted form by many simulation suites including \texttt{Illustris} and \texttt{IllustrisTNG}, temporarily decouples `kicked' particles from the hydrodynamics scheme, in principle aiding numerical convergence (see Section \ref{sec:convergence}) but precluding interactions between winds and the ISM \citep[][]{DallaVecchia2008}.

The physical mechanism by which AGN feedback couples to the ISM remains poorly understood, with a number of channels being plausible, such as radiation pressure on free electrons and/or dust grains, or very high velocity jets. Despite this absence of consensus, there is ample observational evidence to indicate that AGN drive large-scale, high-velocity outflows of ionized and molecular gas \citep[e.g.][]{Maiolino2012,Cicone2014}. Subgrid AGN feedback models generally assume that some fraction, $\epsilon_{\rm f}$, of the radiated luminosity of an SMBH accretion disc couples to the surrounding ISM: the AGN feedback energy $E_{\rm AGN} = \epsilon_{\rm f}\epsilon_{\rm r}\dot{M}_{\rm BH}c^2$, where $\dot{M}_{\rm BH}$ is the accretion rate and $c$ the speed of light \citep[e.g.][]{Springel2005_quasars}. As with stellar feedback, the injected energy can be coupled to the numerically-modelled gas fluid in thermal and/or kinetic form. The simplest approach assumes that the coupling efficiency is fixed, such that the energy injection rate is proportional to the accretion rate. However, motivated by the observation that high-velocity jets are more typically associated with accretion onto SMBHs at small fractions of the Eddington rate \citep{Churazov2005,Merloni2008,Best2012}, some models adopt a relatively low (high) value of $\epsilon_{\rm f}$ when the SMBH is accreting at a high (low) fraction of the Eddington rate \citep[e.g.][]{Sijacki2007,Dubois2012,Weinberger2017,Dave2019}. Since the Eddington rate depends linearly on the mass of the SMBH, this approach can act like a switch, whereby efficient feedback is triggered once the SMBH reaches a threshold mass.

\subsection{Calibration of subgrid feedback models}
\label{sec:subgrid_calibration}

The numerical values of the parameters governing subgrid models can, in some cases, be chosen by reference to observables or theoretical arguments. In general though, the appropriate values are not known a priori and may be resolution dependent. The efficiencies of feedback processes are the most salient example: the microphysics are not well understood theoretically, and observations do not authoritatively characterise outflow properties on the scales at which subgrid models recouple to the hydrodynamics scheme \citep[see e.g.][]{Chisholm2016a,Chisholm2016b}. Moreover, changing the resolution of a simulation often changes the frequency of individual feedback events, and the energy, mass and momentum they each inject. Even if over some timescale the same total energy (for example) is injected, the macroscopic efficiency of feedback can depend on this intermittency because the numerical losses will differ. 

It is therefore increasingly common practice to calibrate the parameters of subgrid feedback models, so that simulations broadly reproduce key properties of the galaxy population. Depending on the sensitivity of the subgrid models to resolution, the parameters may require recalibration if the resolution is changed. The choice of calibration diagnostics is somewhat arbitrary, but clearly they should be well-characterised by observations on scales resolved by the simulations, and should be sensitive to the parameter(s) requiring calibration. Most often, the stellar mass of galaxies is used (either via the observed GSMF, or the inferred stellar-to-halo mass ratio) to calibrate the efficiency of feedback associated with stellar evolution, and the mass of central SMBHs at fixed galaxy stellar mass to calibrate the efficiency of feedback associated with their growth. Reproduction of these quantities alone does not guarantee a realistic galaxy population, so complementary observables may also be used. For example, by also considering the sizes of disc galaxies, the \texttt{EAGLE} simulations more accurately reproduced many important galaxy scaling relations \citep{Crain2015}. \texttt{IllustrisTNG} further considers the cosmic SFR density and halo gas fractions. \texttt{SIMBA}'s feedback efficiencies were tuned only against the GSMF and the $M_\mathrm{BH}-M_\star$ relation, the latter via the accretion efficiency rather than the AGN feedback efficiency. \texttt{Horizon-AGN}'s stellar feedback efficiency was not explicitly calibrated but inferred from the Starburst99 \citep{Leitherer1999} spectrophotometric model. Its AGN feedback model was, in common with \texttt{EAGLE} and \texttt{IllustrisTNG}, calibrated with reference to the $M_\mathrm{BH}-M_\star$ relation. As noted in Section~\ref{sec:intro}, care must be taken not to interpret, as predictions, emergent properties that are closely related to the calibration diagnostics, nor those properties significantly compromised by the simplifications of the subgrid modelling.

\begin{figure}[h]
\includegraphics[width=6.0in]{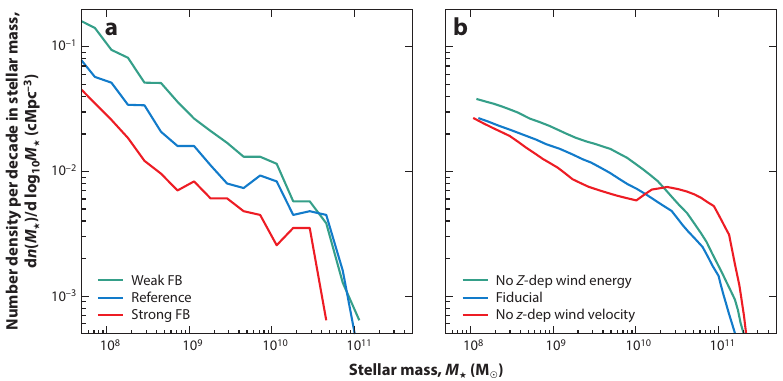}
\caption{An illustration of the influence on the GSMF of adjusting the subgrid stellar feedback model in calibrated simulations. The left panel, adapted from \citet{Crain2015}, shows the response of the present-day GSMF of \texttt{EAGLE} realised in an $L=25\cMpc$ volume to changing the adopted stellar feedback efficiency (which scales linearly with the injected energy) by factors of 0.5 (WeakFB, green curve) and 2.0 (StrongFB, red curve). The right panel, adapted from \citet{Pillepich2018a}, shows the response of the present-day GSMF of \texttt{IllustrisTNG} realised in an $L=37\cMpc$ volume to disabling the metallicity dependence of the injected energy (No Z-dep wind energy, green curve) and to disabling the redshift dependence of the wind injection velocity (No z-dep wind velocity, red curve).}
\label{fig:fb_calib}
\end{figure}

Figure~\ref{fig:fb_calib} shows the influence on the GSMF of adjusting the subgrid stellar feedback model in calibrated simulations. The left panel shows the influence of using stellar feedback efficiencies (effectively a coefficient specifying the fraction of $E_{\rm SN}$ that couples to the ISM) for \texttt{EAGLE}'s stochastic thermal heating treatment, that differ from the Reference model by factors of 0.5 (weak feedback, WeakFB) and 2 (strong feedback, StrongFB). The shifts follow from galaxies of a fixed stellar mass becoming associated with less (more) massive dark matter haloes in the WeakFB (StrongFB) model. Because the dark matter halo mass function is steep, even a small change to the stellar -- halo mass relation significantly alters the space density of galaxies at fixed stellar mass. The right panel shows the influence in \texttt{IllustrisTNG} of disabling the scaling of the energy of its kinetically-driven stellar winds as a function of metallicity (No Z-dep wind energy), and separately, the scaling of the velocity of these winds as a function of redshift (No z-dep wind velocity). The energy scaling (effectively a mass loading scaling at fixed velocity) suppresses the growth of low-mass galaxies, and acts in a fashion similar to that of the \texttt{EAGLE} feedback efficiency scaling in the left panel. The velocity scaling prevents kinetically-driven winds from stalling in relatively massive galaxies \citep[see also][]{Crain2009,Oppenheimer2010}.

In analogy to calibrating the stellar feedback efficiency, a similar procedure is often used to calibrate the efficiency of AGN feedback, whereby the value of $\epsilon_{\rm f}$ is adjusted to achieve broad reproduction of the observed relation between the stellar mass of galaxies and the mass of their central SMBH. \citet{Booth2010} show that so long as AGN feedback is numerically efficient and well sampled, stellar masses are insensitive to extreme variations of $\epsilon_{\rm f}$, as SMBHs compensate by growing to the mass that enables them to inject the feedback energy needed to (self-)regulate gas inflow. As such, $\epsilon_{\rm f}$ primarily sets the normalisation of the SMBH -- galaxy mass scaling relation.

\subsection{Verification and convergence}
\label{sec:convergence}

Numerical simulations, irrespective of their purpose or application, are simplified approximations of real phenomena or systems. It is therefore necessary to verify their outcomes to ensure they are fit for purpose. For many terrestrial applications, simulations can be validated against experimental data (e.g. computational fluid dynamics simulations are often confronted with wind tunnel measurements). Clearly, cosmological simulations can appeal to no such testbed. The performance of hydrodynamics solvers can be examined using a set of idealised tests (e.g.\ shock tubes, point explosions, vortex problems, fluid instability tests), but being a strongly non-linear process resulting from the interplay of complex physical processes, galaxy formation offers no simple tests with analytic or known solutions that can be used for validation. A laborious but illuminating strategy is therefore to apply multiple models to the same problem, enabling an assessment of the consensus between the models \citep[e.g.][]{Scannapieco2012,Kim2014,Cui2018}.

Predictive power demands that the outcomes of simulations are robust to changes of various aspects, such as changes in resolution and the size of the simulation domain, a characteristic often referred to as `convergence'. Unhelpfully, the physics of galaxy formation does not readily lend itself to achieving converged results. As noted in Section \ref{sec:IC}, changes to the size of the simulation volume influence the diversity of the environments and halo population that can be realised by the simulations, and the mass and spatial scales for which cosmic variance becomes important. They can also impact volume-integrated properties, such as the global star formation rate density.

Resolution convergence is particularly challenging to achieve, owing to the dominant role played by physical processes taking place at, or below, the numerical resolution scale (see Figure~\ref{fig:scales}). In simple terms, a higher resolution simulation can resolve smaller galaxy/halo progenitors, enabling them to be followed at earlier cosmic epochs. They can also resolve higher gas densities, influencing (net) radiative cooling rates and, in some cases, interfacing with subgrid models at a different spatial scale. As remarked in Section \ref{sec:subgrid_feedback}, subgrid models can be designed to mitigate this resolution sensitivity, for instance by temporarily decoupling gas from hydrodynamical forces or radiative cooling, and/or by using the generally better-converged properties of the local dark matter distribution, rather than those of the gas, as inputs. However, these choices introduce their own drawbacks, moving the modelling philosophy closer to the phenomenological approach of semi-analytic modelling. As noted in Section \ref{sec:subgrid_calibration}, changing the resolution can also incur more subtle consequences, such as altering the intermittency and numerical efficiency of feedback events. 

It is therefore good practice to compare the properties of galaxy populations that emerge from simulations when the volume (or boxsize) and resolution are varied (ideally individually). Such tests are straightforward and inexpensive to conduct when using smaller volumes and lower resolutions. To mitigate the computational cost of testing for convergence at higher resolution, it is common to conduct high-resolution simulations in a smaller simulation volume, and to run a partner simulation of the same volume at the fiducial resolution to control for boxsize effects. With such a suite of simulations, it is possible to make authoritative assessments of the degree to which the properties of the galaxy population are robust to volume and resolution changes. It is however important to remain vigilant to the possibility that the appearance of a result being converged may be contingent on the particular implementation of one or more subgrid models. 

Many subgrid models employ stochastic treatments of continuous processes, such as the conversion of gas particles or cells into stellar particles, or injecting energy kinetically and/or thermally into the gas through feedback processes. A number of recent studies have examined the impact of this subgrid stochasticity on emergent macroscopic properties, in simulations of individual galaxies \citep[][]{Keller2018,Davies2021,Davies2022} and periodic volumes \citep{Genel2019,Borrow2022b}, finding that small differences due to stochasticity can propagate into significant differences in present-day properties, even for galaxies resolved with many thousands of particles. The systematic uncertainty due to stochasticity is more significant in smaller samples of galaxies, i.e. zooms and small periodic volumes, and needs to be considered when interpreting results, including when analysing small simulations for the purposes of model calibration. In such cases, the uncertainty can be mitigated with brute force, by running multiple realisations of the same simulation using different random number seeds.

\section{Key properties of simulated galaxy populations}
\label{sec:key_properties}

In this section we examine the degree to which state-of-the-art hydrodynamical simulations reproduce key properties of the galaxy population. Clearly the definition of `key' is somewhat arbitrary, and opinions will differ between practitioners, but we focus here on diagnostics that are sufficiently important to be widely considered as validation tests. Marked failure of a simulation to reproduce these diagnostics will undermine confidence in the broader conclusions one may draw from it. In some cases these properties are used as calibration diagnostics, so should not be considered predictions.

\subsection{Galaxy stellar mass function} 
\label{sec:GSMF}

\begin{figure}[h]
\includegraphics[width=4.0in]{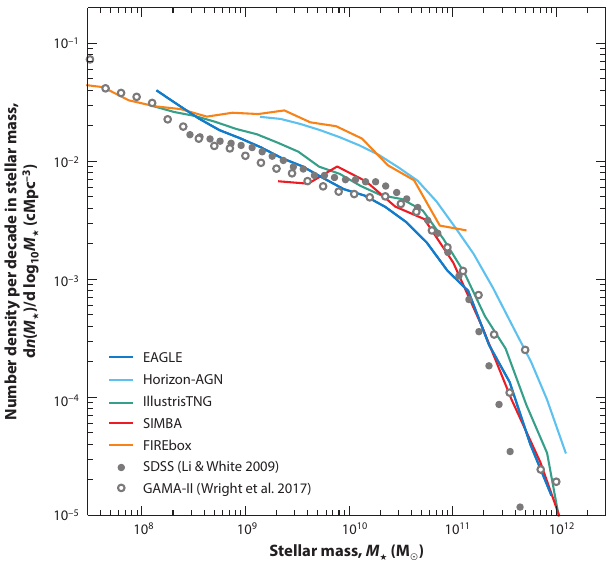}
\caption{The present-day GSMF of the four simulations listed in Table \ref{tab1}: \texttt{EAGLE} (data from \citealt{Schaye2015}), \texttt{Horizon-AGN} (data from \citealt{Kaviraj2017}), \texttt{IllustrisTNG} (data from \citealt{Pakmor2023}), \texttt{SIMBA} (data from \citealt{Dave2019}). We also show the GSMF of the \texttt{FIREbox} simulation (data from \citealt{Feldmann2023}), whose resolution is higher, and simulation volume smaller, than the other simulations. Filled and open black symbols show GSMF measurements from the SDSS and GAMA surveys \citep{Li2009, Wright2017}; for clarity we omit their error bars. \texttt{EAGLE}, \texttt{IllustrisTNG} and \texttt{SIMBA} recover the observed space density of galaxies at fixed mass (over the mass range for which the observational measurements concur) to within $0.3\,{\rm dex}$. Due to overcooling, \texttt{Horizon-AGN} and \texttt{FIREbox} yield a space density for galaxies of $10^{9}-10^{10}\,\Msun$ that is too high by up to $0.5\,{\rm dex}$.}
\label{fig:GSMF}
\end{figure}

The $z\simeq 0$ GSMF is the observational diagnostic most frequently used to calibrate the parameters of subgrid feedback models. Simulations of $L\sim 100\cMpc$ and baryonic mass resolution $m_{\rm g} \sim 10^6\,\Msun$ adequately resolve and sample the GSMF over the range $8 \lesssim \log_{10} (M_\star/\Msun) \lesssim 11$, thus demanding realistic stellar masses for galaxies with space densities that differ by at least two decades. The present-day GSMFs of \texttt{EAGLE}, \texttt{Horizon-AGN}, \texttt{IllustrisTNG} and \texttt{SIMBA} are shown in Figure~\ref{fig:GSMF} together with meaurements from the Sloan Digital Sky Survey (SDSS) and Galaxy and Mass Assembly (GAMA) surveys. This highlights that it has proven possible to reproduce the GSMF over much of this stellar mass range in simulations of the galaxy population, in some cases with an accuracy better than or comparable to the $\sim 0.3\,{\rm dex}$ systematic uncertainty on stellar masses inferred using population synthesis models \citep[e.g.][]{Conroy2009}. Although the \texttt{EAGLE}, \texttt{IllustrisTNG} and \texttt{SIMBA} simulations were calibrated to achieve this reproduction, success was not guaranteed because their subgrid models afford only limited freedom. For comparison, the figure also shows the GSMF of the flagship $L=22.1\cMpc$ \texttt{FIREbox} simulation, in which the parameters governing stellar feedback processes (this simulation does not model AGN feedback) were not calibrated against properties of the galaxy population. \texttt{FIREbox} and \texttt{Horizon-AGN} overproduce the space density of galaxies, especially for those with stellar mass $9 \lesssim \log_{10} (M_\star/\Msun) \lesssim 10$.

A common aspect of models with realistic GSMFs is the operation of efficient feedback at all mass scales. Simulations that do not reproduce the observed GSMF commonly \citep[but not exclusively, e.g.][]{Pakmor2023} form too many galaxies at fixed stellar mass. As noted in Section \ref{sec:subgrid_calibration}, this is a consequence of feedback failing to adequately regulate galaxy growth and thus allowing galaxies to form in low-mass dark matter haloes with too high space density. Failure to regulate galaxy growth can be due to the implemented feedback being unintentionally inefficient for numerical reasons (e.g.\ poor sampling of energy injection events, or injection of too little energy per feedback event), because a feedback mechanism (e.g.\ AGN) is not included, or because a comprehensive search of the model's plausible many-dimensional parameter space is too computationally expensive. The latter is a major barrier to exhaustive calibration, and has motivated the adoption of Gaussian process emulation to accelerate the procedure \citep{Kugel2022}.

Reproduction of the GSMF at $z\simeq 0$ alone is insufficient to ensure that simulations also recover its evolution with cosmic time, because plausible feedback models can yield unrealistic star formation histories (SFHs) \citep[][]{Crain2015}. However, the use of complementary calibration diagnostics (e.g.\ galaxy sizes, or the cosmic SFR density) enables SFHs to be constrained sufficiently to reproduce GSMFs as early as $z\simeq7$, to an accuracy consistent with the systematic uncertainties on observationally-inferred masses \citep[e.g.][]{Furlong2015,Pillepich2018b,Dave2019}. Simulations of $L\sim 100\cMpc$ are ill-suited for confrontation with the demographics of very high-redshift galaxies derived from early \textit{James Webb Space Telescope} (JWST) imaging \citep[e.g.][]{Labbe2023}, as they do not sample the rare fluctuations in the density field that likely seed the formation of the bright, rare sources preferentially detected by these observations \citep[][]{Kannan2023}. 

\begin{marginnote}[]
\entry{SFH}{Star formation history}
\end{marginnote}

\subsection{Size and morphology}

Reproduction of galaxies with realistic sizes has long been viewed as a prominent challenge for cosmological simulations, owing primarily to the recognition that overcooling can lead to spurious angular momentum transfer from cold gas and stars to dark matter \citep{Katz1991}, and thus to the formation of galaxies that are too compact as well as too massive. Clearly, ensuring that galaxies form with broadly realistic stellar masses is then a necessary condition for reproduction of realistic sizes. 

However, the original \texttt{Illustris} simulation yielded present-day galaxies too \textit{large} by a factor of $\simeq 2$ \citep{Snyder2015}, highlighting that realistic masses are not a sufficient condition for realistic sizes. Using \texttt{EAGLE}, \citet{Crain2015} demonstrated that the overcooling and angular momentum problems are somewhat separable: simulated galaxies must not only form a realistic mass of stars, their SFH must also be broadly realistic, so that stars form from natal gas with the correct angular momentum distribution. For this reason, the sizes of present-day disc galaxies were explicitly considered when calibrating the parameters of \texttt{EAGLE}'s subgrid stellar feedback model. Sizes were also considered by \texttt{IllustrisTNG} during the development of their model of stellar winds.

\begin{figure}[h]
\includegraphics[width=6.0in]{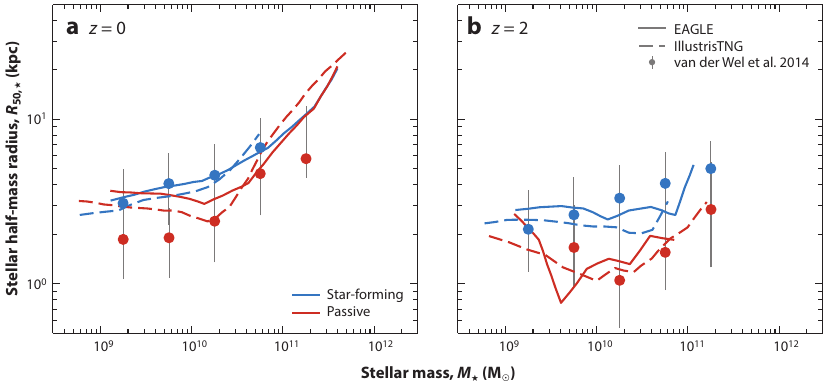}
\caption{Present-day galaxy sizes as a function of stellar mass for star-forming (blue curves) and passive (red curves) galaxies in \texttt{EAGLE} \citep[solid curves, data from][]{Furlong2017} and \texttt{IllustrisTNG} \citep[dashed curves, data from][]{Genel2018}, at $z=0$ (left panel) and $z=2$ (right panel). Filled symbols with error bars denote the observational size measurements of \citet{vanderWel2014}. Both simulations broadly reproduce the observed size -- mass relations of both galaxy types, and their evolution over $\simeq 10\,\Gyr$, though at late times both simulations overestimate the size of low-mass passive galaxies.}
\label{fig:sizes}
\end{figure}

Figure~\ref{fig:sizes} shows the galaxy size -- stellar mass relations of, separately, star-forming and passive galaxies in \texttt{EAGLE} \citep[from][]{Furlong2017} and \texttt{IllustrisTNG} \citep[from][]{Genel2018}, at $z=0$ (panel a) and $z=2$ (panel b). In both simulations, galaxy size is defined as the stellar half-mass radius, $R_{50}$. Filled symbols with error bars are the observational size measurements of galaxies at the corresponding redshifts, from \citet{vanderWel2014}. \texttt{EAGLE} and \texttt{IllustrisTNG} broadly reproduce the observed size -- mass relations of star-forming and passive galaxies, and the evolution of these relations, from smaller sizes at fixed stellar mass, from $z=2$. The simulations differ from the observations by $\simeq 0.1\,{\rm dex}$ at low redshift and $\simeq 0.3\,{\rm dex}$ at early times, and thus compare much more favourably with the observations than was the case for prior simulations \citep[e.g.][]{McCarthy2012}.  

Interestingly, both \texttt{Horizon-AGN} \citep[see][]{Dubois2016} and \texttt{SIMBA} \citep{Dave2019} exhibit size -- mass relations for which, at fixed mass, passive galaxies are larger than star-forming counterparts. In both cases, the relation for star-forming galaxies is in broad agreement with observational measurements, indicating that their passive galaxies are too large. Although in \texttt{EAGLE} and \texttt{IllustrisTNG} star-forming galaxies tend to be larger than passive counterparts, as is observed, the sizes of both galaxy types are similar at low mass (likely due to the poor sampling of SFHs), indicating that passive galaxies in these simulations are also too extended. This may be a consequence of spurious and prolonged dynamical heating of their stellar particles by more massive dark matter particles \citep{Ludlow2019}; if so, this problem can be mitigated via the adoption of baryonic and dark matter particles of similar mass \citep[e.g. as per][]{Tremmel2015}, at the expense of a greater computational cost and memory footprint.

An outcome seen in both \texttt{EAGLE} and \texttt{IllustrisTNG} is a correlation, at fixed stellar mass, between the size and the specific SFR of galaxies. Interestingly, in \texttt{IllustrisTNG} the trend is not seen when one considers only star-forming galaxies, and the overall trend reflects the changing relative fractions of (extended) star-forming and (compact) passive galaxies, from low to high stellar mass. In contrast, massive ($M_\star \gtrsim 10^{9.5}\,\Msun$) star-forming galaxies in \texttt{EAGLE} exhibit a clear correlation between size and specific SFR. 

A related issue is whether simulated galaxy populations reproduce the diversity of observed morphologies (and kinematics). \texttt{EAGLE}, \texttt{Horizon-AGN}, and \texttt{IllustrisTNG} reproduce galaxies with the characteristic early- and late-type morphologies of the Hubble Sequence \citep[see][]{Dubois2014,Schaye2015,HuertasCompany2019}, and the correspondence of morphology (and/or kinematics) with the position of galaxies in the colour -- magnitude diagram or its proxies \citep[][]{Correa2017}. This is not an outcome that the simulations were (or could meaningfully be) calibrated to reproduce and, in common with sizes, largely follows from the build up of angular momentum in intergalactic gas due to large-scale tidal torques, the dynamical coherence of gas as it is accreted onto haloes and galaxies, the preferential removal of low-angular momentum gas in feedback-driven outflows, and the transfer of angular momentum from stars to the dark matter as discs are disrupted by gravitational instabilities and mergers. Detailed analysis of the simulations reveals correlations between the morphology of galaxies and properties of their host haloes, such as angular momentum \citep[e.g.][]{Zavala2016,Yang2021}, flattening \citep{Thob2019}, and assembly time \citep{Davies2020,Davies2021}. However, it should be borne in mind that the simulations can reproduce observed morphologies only in a relatively broad sense, owing to their internal structure being too smooth and their vertical scale heights too large (see Section \ref{sec:subgrid_ism}). It will be fascinating to assess the degree to which future simulations, incorporating detailed models of the multiphase ISM, are able to recover the detailed internal structure and kinematics of the galaxy population.

\subsection{Galaxy clustering}

Accurate reproduction of observed clustering statistics, which connect galaxies to the properties of their dark matter environment, represents an important validation test of galaxy population simulations, particularly because clustering properties are not used as calibration diagnostics. Authoritative predictions also provide a means to stress-test (semi-)analytic methodologies \citep{ChavesMontero2016,Guo2016}, widely used to generate predictions for the large, mildly non-linear scales probed by cosmological surveys.

\citet{Springel2018} showed that \texttt{IllustrisTNG} (both TNG100 and its lower resolution, $L=302.6~\cMpc$ counterpart TNG300) broadly reproduces, on scales of $1 \lesssim r \lesssim 10\,h^{-1}\,\Mpc$ and as a function of stellar mass and colour, the clustering of galaxies revealed in the low-redshift cosmos by SDSS observations. Similarly, \citet{Artale2017} examined the clustering of \texttt{EAGLE} galaxies on scales of $1 \lesssim r \lesssim 6\,\Mpc$ as a function of mass and colour, and found it to be broadly consistent with that measured from the GAMA survey, with the exception of low-mass (and poorly resolved) red galaxies, which cluster too strongly. \citet{Crain2017} showed that the clustering of gas-rich galaxies in \texttt{EAGLE} is in excellent agreement with that inferred from 21cm surveys, and that the dependence of clustering on galaxy colour (at fixed stellar mass) also manifests as a dependence on atomic hydrogen fraction. 

\subsection{Star formation histories}

The evolution of the cosmic SFR density (i.e.\ the volumetric SFR) is a fundamental observable whose measurement is a forefront goal in observational astronomy. Its precise characterisation is challenging owing to selection effects and observational systematic uncertainties \citep[see e.g.][]{Madau2014,Behroozi2019}, but it remains a natural validation benchmark for simulations of representative cosmic volumes. 

\begin{figure}[h]
\includegraphics[width=5.0in]{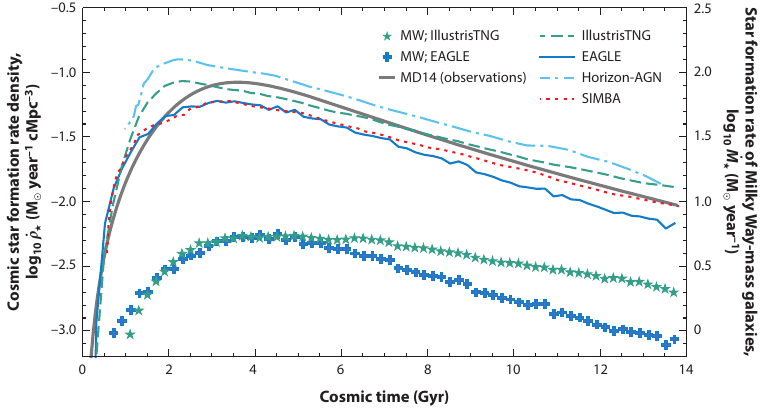}
\caption{The coloured curves show the cosmic SFR density history of \texttt{EAGLE}, \texttt{Horizon-AGN} (adapted from \citealt{Kaviraj2017}), \texttt{IllustrisTNG} and \texttt{SIMBA} (adapted from \citealt{Dave2019}) (colours per legend). The solid black curve denotes the fit to observations compiled by \citet{Madau2014}, converted to the \citet{Chabrier2003} IMF. Coloured symbols show the typical SFH (per the right y-axis) in \texttt{EAGLE} (crosses) and \texttt{IllustrisTNG} (stars) of central galaxies hosted by present-day haloes with mass $M_{\rm halo} \simeq 10^{12}\,\Msun$, broadly similar to that of the Milky Way. The cosmic SFR densities in modern simulations agree with each other to within a factor of a few and the shapes are fairly similar, decreasing substantially over the last 10 Gyr. However, the star formation history of a subset of galaxies can show larger variations as shown by the average SFHs of central galaxies in Milky Way-mass haloes, which diverges towards the present day for \texttt{EAGLE} and \texttt{IllustrisTNG}.}
\label{fig:SFRD}
\end{figure}

Solid, dot-dashed, dashed, and dotted curves in Figure~\ref{fig:SFRD}, corresponding to the left y-axis, show the evolution of the cosmic SFR density of \texttt{EAGLE}, \texttt{Horizon-AGN}, \texttt{IllustrisTNG}, and \texttt{SIMBA}, respectively. The simulations that broadly reproduce the present-day GSMF yield similar SFR density histories (at the factor $\simeq 2$ level). The integral of the cosmic SFR density (over cosmic time) differs from the integral of the GSMF (over mass) only due to mass loss from stellar evolution, and a small correction due to the use of apertures to measure the stellar mass of galaxies. In the context of observational measurements, a factor of $2$ is non-negligible, being comparable to the long-standing tension between the observed cosmic stellar mass density and the mass loss-corrected integral of the cosmic SFR density, though this apparent discrepancy may be resolved by the use of more sophisticated panchromatic spectral energy distribution models \citep{Leja2019}.

The symbols in Figure~\ref{fig:SFRD}, relative to the right y-axis, show the average SFH of central galaxies hosted by present-day haloes with mass $M_{\rm halo} \simeq 10^{12}\,\Msun$, broadly similar to the halo mass of the Milky Way, in \texttt{EAGLE} (crosses) and \texttt{IllustrisTNG} (stars). These histories peak at similar epochs with similar SFRs, but diverge for $t \gtrsim 5\,\Gyr$ ($z \lesssim 1.3$), declining more in \texttt{EAGLE} than in \texttt{IllustrisTNG}. Clearly, model-to-model differences at a particular halo mass scale are not simple rescalings of the difference in cosmic SFR density, and instead reflect the complicated mass dependence of the physical processes treated by the simulations. This mass dependence is also manifest as subtle but significant differences in the shape of the GSMF realised by the simulations, such that the decline of the typical Milky Way-like SFH in \texttt{EAGLE} is likely related to the simulation `undershooting' the space density of galaxies at the knee of the GSMF. The SFHs of galaxies in simulations of the galaxy population are also impacted by the simplified subgrid treatment of the ISM (Section \ref{sec:subgrid_ism}), which yields artificially smooth gas distributions and, by extension, artificially smooth SFHs. This hinders the use of SFHs from the simulation to interpret observations \citep[e.g.][]{Sparre2017} and may have significant consequences for galaxy evolution \citep[][]{Pontzen2012}.

\subsection{Galaxy colours} 
\label{sec:colours}

Galaxy surveys have revealed a remarkable bimodality in the colour distribution of galaxies. Star-forming galaxies appear blue, whereas passive (or `quiescent') galaxies are redder, owing to their lack of short-lived blue stars. There remains much ongoing debate concerning which processes regulate and quench star formation, with such processes usually categorised as internal or external/environmental. We focus here on colour bimodality in relatively massive (and predominantly central) galaxies, and defer discussion of the evolution of satellite galaxies embedded in larger host haloes to Section \ref{sec:satellites}.

Semi-analytic models first highlighted that the emergence of colour bimodality is a natural consequence of the use of energetic feedback in massive galaxies as a means to shape the high-mass end of the present-day GSMF away from the power-law distribution of dark matter haloes \citep[e.g.][]{Kang2005}. To date, this shaping is achieved in all realistic state-of-the-art simulations via the onset of efficient AGN feedback as the dominant regulation mechanism in haloes of mass $M_{\rm halo}\gtrsim 10^{12}\,\Msun$. Because these simulations are usually calibrated to reproduce the GSMF, it is not unreasonable to expect that they yield bimodal galaxy colours. However, accurate reproduction of the observed distribution of galaxies in the colour -- mass plane also relies on the simulations yielding realistic SFHs and metallicities, because these properties also influence the observable properties of galaxies and are hence inputs to the population synthesis models used to translate physical properties into observables. One must also apply reasonable corrections for obscuration and attenuation by dust. 

\texttt{EAGLE} \citep{Trayford2015,Trayford2017} and \texttt{IllustrisTNG} \citep{Nelson2018colour} have been shown to reproduce the key features of the colour -- mass plane recovered from galaxy surveys. Having satisfied this validation step, the simulations enable the origin of bimodality to be examined in detail, and the sensitivity of the colour -- mass distribution to details of the models to be explored. \texttt{EAGLE} and \texttt{IllustrisTNG} present the concensus that bimodality is a direct consequence of the onset of AGN feedback. The presence of blue galaxies that are brighter than those observed in \texttt{EAGLE} likely indicates that its AGN feedback does not adequately quench some fraction of massive galaxies \citep{Trayford2015}. Similarly, the greater fraction of red discs and blue spheroids in \texttt{IllustrisTNG} compared to observations may indicate that efficient AGN feedback is triggered in the wrong objects or at the wrong time \citep{RodriguezGomez2019}; this is a problem whose resolution may require that AGN triggering is more sensitive to mergers \citep{Bustamante2019}. \citet{Donnari2021} compare the closely related fraction of quenched galaxies as a function of stellar mass for a number of state-of-the-art simulations, finding qualitative but not quantitative agreement for central galaxies.

\section{Galaxy scaling relations}
\label{sec:scaling_relns}

We next examine the degree to which well-known galaxy scaling relations are reproduced by state-of-the-art simulations of the galaxy population. The form of, and scatter about, these scaling relations encodes valuable information concerning the physics of galaxy evolution, but extracting insight from observed relations often requires the interpretive power of simulations. Confidence in the insight obtained is naturally greater if multiple simulations present a consensus for the origin of particular relations. As with Section \ref{sec:key_properties}, we caution that some of the scalings presented here are popularly used as calibration diagnostics, and cannot be treated as predictions. 

\subsection{Supermassive black holes}
\label{sec:scalings_SMBH}

The primacy of AGN feedback as the mechanism by which state-of-the-art simulations regulate and quench star formation in massive galaxies, renders the relationship between the mass of galaxies and that of their central SMBH of particular importance. Although AGN feedback is not thought to dominate in low-mass galaxies, the latter are the sites in which SMBHs are seeded. SMBHs grow in concert with their host galaxy, yielding a power-law scaling between their masses. Whether this relation is causal is the subject of energetic debate, as it may for example reflect a natural outcome of hierarchical assembly \citep{Jahnke2011}, though it has been argued that this explanation is difficult to reconcile with the masses of SMBHs in dwarf galaxies \citep{King2021}.

\begin{figure}[h]
\includegraphics[width=5.0in]{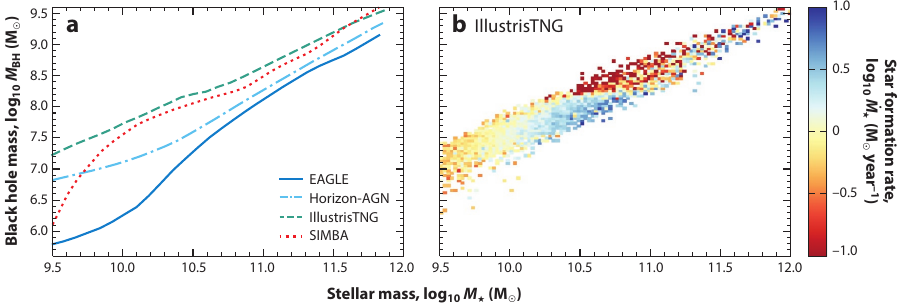}
\caption{Left: the characteristic present-day SMBH mass ($M_\mathrm{BH}$) as a function of stellar mass ($M_\star$) for \texttt{EAGLE}, \texttt{Horizon-AGN}, \texttt{IllustrisTNG} and \texttt{SIMBA}, adapted from \citet{Habouzit2021} Right: as left, but showing the distribution of present-day central galaxies in \texttt{IllustrisTNG}, coloured by their average SFR. For galaxies with $10^{10}\lesssim M_\star\lesssim10^{11}\,\Msun$, the SFR is lower in galaxies with more massive black holes, because they have experienced stronger AGN-driven outflows.}
\label{fig:BH}
\end{figure}

The left panel of Figure~\ref{fig:BH}, adapted from \citet{Habouzit2021}, shows the relation between the mass of the SMBHs and their host galaxies in \texttt{EAGLE}, \texttt{Horizon-AGN}, \texttt{IllustrisTNG} and \texttt{SIMBA}. The four simulations are broadly consistent with observed SMBH masses for $M_\star \gtrsim 10^{10.5}\,\Msun$. In this regime the simulations exhibit scaling relations with similar slope and, as remarked in Section \ref{sec:subgrid_calibration}, the normalisation of the relation largely reflects differences in the calibration/choice of the subgrid AGN efficiency parameter, $\epsilon_{\rm f}$. There is poorer consensus at lower galaxy masses, a regime in which the scatter in observed masses is also large  \citep[e.g.][]{Kormendy2013}. In this regime, the growth of SMBHs is likely regulated by stellar feedback \citep[e.g.][]{Dubois2015,Bower2017}, and the growth of simulated SMBHs is sensitive to degenerate details such as the seed mass, the subgrid accretion model, and subgrid treatments of unresolved dynamics \citep[e.g.][]{Bahe2022}. 

The right panel of Figure~\ref{fig:BH} shows the present-day SMBH scaling relation in more detail for \texttt{IllustrisTNG}, with symbols coloured by the SFR of the host galaxy. In massive galaxies ($M_\star \gtrsim 10^{10}\Msun$), there is a strong negative correlation, at fixed stellar mass, between SMBH mass and SFR, indicative of the regulation of galaxy SFRs by AGN-driven outflows. Importantly, there is a qualitative consensus in state-of-the-art simulations for this particular correlation. However, as shown by \citet[][see their Fig 4]{Habouzit2021}, the simulations differ significantly in terms of the quantitative influence of AGN feedback on star formation regulation.

\subsection{The star-forming main sequence} 
\label{sec:scalings_mainsequence}

Star-forming galaxies exhibit a tight correlation between their SFR and stellar mass; this relation has become popularly known as the `star-forming main sequence' \citep[SFMS, e.g.][]{Noeske2007}, whose normalisation increases with increasing redshift. The ubiquity of the relation has led to the adoption of a more refined definition of what constitutes a passive galaxy, shifting from the canonical threshold at $\dot{M}_\star > 10^{-11}$~yr$^{-1}$ to some number of decades from the median of the SFMS, thus naturally accounting for the mass and redshift dependence of the characteristic SFR of star-forming galaxies.

State-of-the-art simulations broadly reproduce the present-day SFMS, elucidating its origin and that of the scatter about it. Analysing the SFMS in \texttt{IllustrisTNG}, \citet{Donnari2019} found that its normalisation is mildly sensitive to both the timescale over, and aperture within which, the SFR measured, and that its intrinsic scatter (roughly at the factor of two level) is largely insensitive to stellar mass. However they caution that the redshift evolution of the SFMS is not as strong as inferred from observations, a shortcoming in common with prior generations of simulations \citep[e.g.][]{Dave2008}. \citet{Matthee2019} examined the SFMS in \texttt{EAGLE}, finding a mild mass dependence. They showed that scatter in the present-day SFMS is due to both fluctuations on short time scales ($\lesssim 2\Gyr$) associated with the physics of self-regulated gas flows, and long time scale ($\sim 10\Gyr$) variations due to differences in halo formation histories. Related consequences of the short duration fluctuations are correlations at fixed stellar mass between the SFR and the outflow velocity of winds \citep{Nelson2019}, and between the SFR and the mass of cold gas that is (or will soon become) available for star formation \citep[e.g.][]{Lagos2016,Appleby2020}, the latter being readily corroborated by observations \citep{Saintonge2022}. 

\subsection{The Tully-Fisher relation}
\label{sec:scalings_tullyfisher}

The \citet{Tully1977} relation is a well-known scaling relation connecting the asymptotic rotation velocity of disc-dominated galaxies and their luminosity (or mass). The strongest correlation is recovered when considering the total baryonic mass of galaxies \citep{McGaugh2000}. A similar relation exists for elliptical galaxies, between their central velocity dispersion and their mass (or luminosity; \citealt{Faber1976}). These relations are the traditional means by which the properties of galaxies were linked to those of their haloes, because galaxy dynamics (in standard cold dark matter cosmogonies) are primarily governed by the structure of their haloes. 

The failure of prior generations of simulations to reproduce the observed rotation profiles of individual galaxies \citep[summarised succinctly by][]{Scannapieco2012}, largely as a consequence of overcooling, translated in to a failure of simulations of galaxy populations to reproduce the high-mass end of the Tully-Fisher relation \citep[e.g.][]{McCarthy2012}. The inclusion of AGN feedback and the calibration of feedback models in state-of-the-art simulations results in a more realistic stellar mass -- halo mass relation and prevents artificial contraction of the halo due to the  excessive condensation of stars \citep[e.g.][]{Schaller2015}. As a result, these simulations broadly reproduce the observed Tully-Fisher relation \citep[e.g.][]{Vogelsberger2014mn, Ferrero2017, Sales2017}, without having been explicitly calibrated to do so.

\subsection{The mass -- metallicity relations}
\label{sec:scalings_massmetallicity}

Feedback-driven outflows transport a fraction of the heavy elements synthesised in galaxies into the CGM \citep{Tumlinson2011}, and even beyond to the IGM \citep{Aguirre2001}. This establishes a relationship between the mass of a galaxy and the characteristic metallicities of its gas and stars \citep{Larson1974}. This `mass -- metallicity relation' was revealed in detail by the advent of highly-multiplexed spectroscopic surveys such as SDSS, which enabled both the gas-phase metallicity \citep{Tremonti2004}, and that of the stars \citep{Gallazzi2005}, to be measured for 10s-100s of thousands of galaxies. These studies thus reveal that a greater fraction of the metals synthesised by low-mass galaxies, relative to more massive counterparts, are transported into their gaseous environments. These trends are broadly reproduced by \texttt{EAGLE} \citep{Schaye2015}, \texttt{Horizon-AGN} \citep{Dubois2014}, \texttt{IllustrisTNG} \citep{Torrey2019} and \texttt{SIMBA} \citep{Dave2019}, and are a natural consequence of mass-dependence of the macroscopic mass loading of outflows that emerges from their subgrid models for feedback \citep[][see also \citealt{Muratov2015,Christensen2016}]{Beckmann2017,Nelson2019,Mitchell2020}. 

The observations reveal significant scatter in metallicity at fixed stellar mass, and other properties have been shown to correlate with this scatter, such as star formation rate (the `fundamental metallicity relation'); this behaviour is reproduced by the simulations \citep[e.g.][]{Lagos2016,Dave2019,Torrey2019}. \citet{DeRossi2017} showed using \texttt{EAGLE} that one might expect a complex mass dependence on the correlation of metallicity with gas fraction and star formation rate. In low mass galaxies, lower metallicities correspond to higher gas fractions, higher star formation rates and the presence of young stellar populations, whereas at higher mass scales lower metallicities correspond to gas poor, quiescent galaxies. They interpreted this inversion as a consequence of the switch from stellar feedback-based regulation in low-mass galaxies, to AGN-dominated regulation.

\subsection{Cold gas in galaxies}
\label{sec:scalings_cold}

Forging an holistic view of galaxy evolution requires an understanding of the evolution of their gaseous components, which can span many decades in density, temperature and ionisation fraction. Recovery of these properties is a prerequisite for the use of the simulations to elucidate the cycling of baryons into, and out of, galaxies \citep[e.g.][]{Tumlinson2017,Peroux2020araa}, which is driven by the interplay of large-scale cosmic structure evolution and the complex physics of galaxy formation. 

Scaling relations relating the mass of cold, dense gas (primarily atomic and molecular hydrogen) can be readily constructed from observations of galaxies in the low-redshift cosmos \citep[see e.g.][]{Saintonge2022}. Such relations present an appealing benchmark, but because large simulations of the galaxy population do not in general employ an explicit radiation transport scheme, they cannot model the balance of molecular, atomic, and ionised hydrogen on-the-fly in a self-consistent fashion. The partitioning is therefore achieved as a post-processing step based on empirical relationships and/or theoretical models. 

Building on prior examinations of \texttt{EAGLE} \citep[][]{Lagos2015,Bahe2016,Crain2017} and \texttt{IllustrisTNG} \citep{Diemer2018,Stevens2019}, \citet{Dave2020} compared the cold gas mass functions and scaling relations of these simulations with those of \texttt{SIMBA}. The comparison revealed a broad qualitative concensus, but also a number of significant quantitative differences, such as in the characteristic break scale of the \HI\ mass function and the direction in which this quantity shifts with increasing redshift. None of the simulations were found to universally reproduce the key cold gas observational diagnostics, with each exhibiting a different conflict with the data, despite all three simulations broadly reproducing the present-day GSMF and SFMS. This signals that the combined action of the subgrid treatments of interstellar gas, star formation, and feedback in these simulations is too simplistic to reproduce holistically the cold gas and stellar properties of present-day galaxies.

\section{Cosmic gas}
\label{sec:cosmic_gas}

The explicit modelling of cosmic gas flows and their interaction with galaxies is a major selling point of hydrodynamical simulations, relative to simplified methods such as semi-analytic modelling. The inclusion of energetic feedback processes in hydrodynamical simulations has indicated that a complex and intimate connection likely exists between the life cycles of galaxies and their gaseous environments, for which there is a growing body of corroborating observational evidence \citep[e.g.][]{Peroux2020araa}. Besides yielding fascinating insight in its own right, study of the gaseous cosmos is therefore now widely acknowledged as a complementary and profitable means of understanding galaxy formation and evolution, particularly in regard to constraining the physics of feedback and metal transport. In this section we review key properties of cosmic gas, on the scale of the cosmic large-scale structure (the IGM) and within galaxy haloes (the CGM/ICM), that emerge from state-of-the-art simulations of the galaxy population. The confrontation of these outcomes with observational measurements represent strong tests of the simulations, because such properties are less frequently considered than stellar properties when calibrating subgrid feedback models.

\subsection{Absorption system statistics}
\label{sec:abs_stats}

The low-redshift column density distribution functions of intermediate- and high-ionisation state metal ions in \texttt{EAGLE} and \texttt{IllustrisTNG} have been presented in a number of studies \citep[e.g.][]{Schaye2015,Nelson2018oxygen,Wijers2019}. These demonstrate that the simulations are mostly compatible with observational measurements of the column density distribution function of absorption systems seen in quasar spectra, which exhibit significant scatter at fixed column density. Differences between the simulations are limited to the relatively high column regime ($N \gtrsim 10^{15}\percmsq$), in which \texttt{EAGLE} yields too few \OVI\ absorbers, and \texttt{IllustrisTNG} yields too many. In both \texttt{EAGLE} and \texttt{IllustrisTNG}, the abundance of \OVI\ absorbers increases slightly with increasing resolution. \citet{Wijers2019} demonstrated that the abundance of high column density \OVII\ and \OVIII\ absorbers is influenced by AGN feedback in a non-trivial fashion: AGN-driven outflows increase the metallicity of absorbers by transporting oxygen into halo outskirts, but reduce the characteristic density of the absorbing gas. However, the close correspondence of the distribution functions of \OVI, \OVII\ and \OVIII\ in both simulations indicates that these diagnostics are, perhaps surprisingly, only mildly sensitive to the choice of hydrodynamics solver and details of how AGN and stellar feedback processes are implemented.  

A related diagnostic is the covering fraction of absorption systems in the vicinity of galaxies. Prior generations of simulations failed to reproduce the high covering fraction of high column density ($N \gtrsim 10^{17}\percmsq$) neutral hydrogen absorbers associated with star-forming galaxies at $z\simeq 2-3$ \citep[e.g.][]{Fumagalli2014}. Using \texttt{EAGLE}, \citet{Rahmati2015} demonstrated that the \HI\ covering fraction is largely determined by the supply of neutral hydrogen into haloes, which is governed primarily by gravitational infall, and only weakly influenced by feedback processes. However, feedback crucially governs the relationship between galaxy stellar mass and halo mass, such that recovering realistic covering fractions for simulated galaxies of a fixed stellar mass requires that they form in realistic environments. The greater feedback efficiency of state-of-the-art simulations relative to earlier generations of simulations ensures that $z\simeq 2-3$ star-forming galaxies form in more massive haloes, which exhibit radial \HI\ covering fraction profiles consistent with those observed. 

\subsection{Physical properties of the circumgalactic medium}
\label{sec:phys_cgm}

The physical properties of the CGM have long been viewed as a powerful means of constraining the physics of galaxy formation, because the CGM interfaces galaxies with their incoming supply of fuel for star formation and their feedback-driven outflows of enriched gas. A salient example is the opportunity to elucidate how outflows operate: the stellar masses of galaxies (for example) are largely agnostic to whether cosmic gas inflows are regulated by violent, intermittent episodes of feedback or a gentler and more continuous process, but the physical conditions of the CGM are expected to differ strongly between these cases \citep[e.g.][]{vandeVoort2012}. 

Using \texttt{EAGLE}, \citet{Oppenheimer2018} examined the physical conditions traced by low-ionisation state absorption systems in the CGM of galaxies with mass comparable to that of the Milky Way, to assist the interpretation of results from the COS (Cosmic Origins Spectrograph)-Halos survey \citep{Werk2013,Werk2014}. They recovered column densities for most ions within a factor of $\simeq 2$ of those reported by COS-Halos. They found little correlation of absorber column densities with the specific SFR of central galaxies, but a significant decline of column densities at larger impact factors, both findings being consistent with the observations. Their analysis elucidated that these low-ionisation state absorption systems trace cool ($T\sim 10^4\,\K$) clumps of gas close to galaxy discs ($\lesssim  100\kpc$), distinct from the warmer ($T\simeq 10^{5.5}\,\K$) and more diffuse gas traced by \OVI\ absorbers, usually found at $\gtrsim 150\kpc$. This suggests that low ionisation state absorbers likely trace enriched gas that is reaccreting onto galaxies.

Perhaps unsurprisingly, comparison of similar CGM diagnostics from different simulations reveals areas of both consensus and discrepancy. \citet{Peroux2020} demonstrate that the gas mass fluxes and metallicities as a function of the azimuthal angle about star-forming galaxies are qualitatively similar in \texttt{EAGLE} and the high-resolution $L=51.7~\cMpc$ TNG50 simulation of the \texttt{IllustrisTNG} suite. Both simulations exhibit a strong azimuthal dependence such that the major axis is dominated by relatively metal-poor inflows and the minor axis by enriched outflows. They also broadly reproduce the dichotomy of \OVI\ column densities associated with star-forming and passive galaxies revealed by COS-Halos \citep{Tumlinson2011}. However, the origin of the relationship differs between the simulations. Based on analysis of \texttt{EAGLE}, \citet{Oppenheimer2016} argue that passive galaxies exhibit a deficit of \OVI\ because they preferentially reside in more massive haloes (in which oxygen tends to be more strongly ionised) than star-forming galaxies of comparable stellar mass. \citet{Nelson2018oxygen} report a more direct connection between star formation and the transport of oxygen-enriched gas into the CGM by AGN feedback in \texttt{IllustrisTNG}. Such differences may be the tip of an iceberg, as the inclusion of physical processes not treated by either of \texttt{EAGLE} or \texttt{IllustrisTNG}, for example feedback by cosmic rays, may induce significant changes to the CGM \citep{Ji2020, Butsky2022}.

\subsection{Gas inflows \& outflows}
\label{sec:gasflows}

\begin{figure}[h]
\includegraphics[width=5.0in]{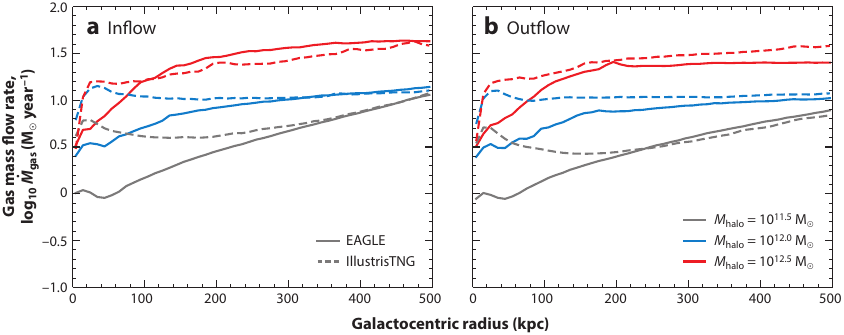}
\caption{Inflow (left panel) and outflow (right panel) rates as a function of galactocentric radius for $M_\mathrm{halo}=10^{11.5}$~M$\odot$ (black curves), $10^{12}$~M$\odot$ (blue curves), and $10^{12.5}$~M$\odot$ (red curves) in \texttt{EAGLE} (solid curves) and \texttt{IllustrisTNG} (dashed curves) at $z=0$. Even though both simulations produce similarly realistic galaxies, the flow of gas into and out of galaxies and their haloes can be substantially different. Additionally, the magnitude and sign of these differences depend on halo mass.}
\label{fig:inout}
\end{figure}

The ISM and CGM are essentially gas reservoirs bound to galaxy haloes, whose mass evolves in response to a shifting balance between, primarily, galaxy- and halo-scale inflows and outflows, with a generally minor contribution also played by star formation, stellar mass loss and, in some cases, external environmental processes. In Figure~\ref{fig:inout} we show the halo mass-dependent present-day inflow and outflow rates as a function of galactocentric radius in \texttt{EAGLE} (solid curves) and \texttt{IllustrisTNG} (dashed curves). Black curves correspond to haloes of mass $10^{11.5}~\Msun$, blue curves to haloes of $10^{12}~\Msun$, and red curves to haloes of $10^{12.5}~\Msun$. Comparison to the present-day SFRs of the central galaxies hosted by $10^{12}~\Msun$ haloes shown in Figure~\ref{fig:SFRD} highlights that the inflow and outflow rates are higher than the SFRs for both simulations. 

Subgrid feedback models aim to approximate the impact of unresolved feedback processes on resolved scales. The flow rates presented in Figure~\ref{fig:inout} illustrate that, despite the subgrid treatments in \texttt{EAGLE} and \texttt{IllustrisTNG} yielding broadly similar present-day galaxy populations, the radial transport of gas through their haloes differs significantly \citep[see also][]{Mitchell2020}. For example, inflow rates in \texttt{IllustrisTNG} haloes are fairly constant as a function of galactocentric radius, whereas those in \texttt{EAGLE} decrease more strongly towards the halo centre. Outflow rates in the two simulations are reasonably similar for the more massive haloes ($10^{12.5}\Msun$) and at large radii, whereas inflow rates are different by a factor of 2 in the low and high halo mass range. Within $\simeq 50\kpc$ of the centre of the less massive haloes, the simulations differ by a factor of 3 or more.

Differing rates of inflow and outflow lead to differing halo baryon fractions, with high (net) inflow rates at the virial radius (as seen for low mass haloes in \texttt{IllustrisTNG}) being conducive to a high baryon fraction (see Figure~\ref{fig:fbar}) if sustained over a significant fraction of cosmic time. In the absence of feedback and large-scale outflows, the gas accretion rate at the virial radius is similar to the product of the cosmic baryon fraction, ($\Omega_{\rm b}/\Omega_{\rm m})$, and the dark matter accretion rate, but large-scale feedback-driven outflows can significantly suppress this accretion rate and so also act as a form of `preventative' feedback \citep[e.g.][]{Wright2020} that inhibits intergalactic gas from reaching the CGM and the ISM. The extent to which this happens is likely to depend strongly on the subgrid implementation of the feedback, and the differing flow rates shown in Figure~\ref{fig:inout} indicate that state-of-the-art simulations do not present a consensus.

\subsection{Halo baryon fractions}
\label{sec:fbar}

\begin{figure}[h]
\includegraphics[width=5in]{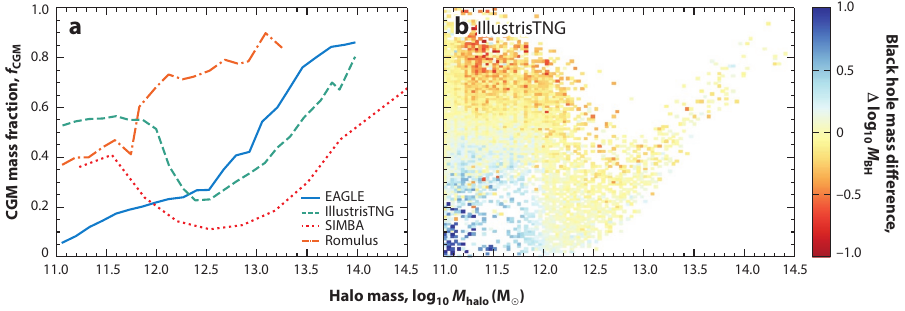}
\caption{The gas fraction in the CGM (the CGM mass divided by the mass of the halo, normalised by the cosmic average baryon fraction) as a function of halo mass. The left panel shows the median for \texttt{EAGLE}, \texttt{IllustrisTNG}, \texttt{SIMBA} (data provided by R.~Dav\'e) and \texttt{Romulus} (data provided by M.~Tremmel). The full distribution for \texttt{IllustrisTNG} is shown in the right panel, colour coded by the average logarithmic difference in SMBH mass as compared to the median SMBH mass at the same halo mass. Haloes with more massive black holes have lower gas fractions for $M_\mathrm{halo}\lesssim10^{12.5}$~M$_\odot$, because these haloes have experienced more feedback from AGN.}
 \label{fig:fbar}
\end{figure}

The baryon fraction of haloes, i.e. the ratio of the baryonic and total masses within the virial radius, is a closely related diagnostic to halo gas flows. In the absence of radiative physics (cooling and feedback), simulations indicate that baryon fractions should be close to the cosmic average baryon fraction of $\Omega_{\rm b}/\Omega_{\rm m} \approx 0.16$ \citep{Crain2007}. Feedback-driven outflows, however, can markedly reduce the baryon fraction. Figure~\ref{fig:fbar} shows the median halo CGM mass fractions as a function of halo mass for \texttt{EAGLE}, \texttt{IllustrisTNG}, \texttt{SIMBA} and \texttt{Romulus} \citep{Tremmel2017}. We include the latter here as it features a markedly different scheme for accretion onto SMBHs. 

The baryon fractions of the massive haloes that host rich galaxy groups and clusters are reasonably well constrained \citep{Vikhlinin2005,Sun2009}, and it is now well established that, owing to differences in the subgrid implementations of feedback processes, accurate reproduction of the GSMF offers no guarantee of accurately recovering realistic baryon fractions on these scales \citep[e.g.][]{Schaye2015,Haider2016}. However, adjustment of the feedback implementations, particularly for AGN, enables the correspondence with observed baryon fractions to be improved \citep{Schaye2015,Weinberger2017,McCarthy2017}. In less massive haloes, gas fractions are largely unconstrained and, as shown in Figure~\ref{fig:fbar}, the median relations that emerge from state-of-the-art simulations are strikingly dissimilar. As first highlighted by \citet{Davies2020}, the median present-day CGM mass fraction at $M_\mathrm{halo}=10^{12}$~M$_\odot$ is $\simeq 0.2$ in \texttt{EAGLE}, but $> 0.5$ in \texttt{IllustrisTNG}.  

Despite the dissimilar trends between CGM mass fraction and halo mass in \texttt{EAGLE} and \texttt{IllustrisTNG}, \citet{Davies2020} showed that in both simulations the scatter in CGM mass at fixed halo mass correlates strongly with the formation redshift of the halo (negatively), the mass of the central galaxy's SMBH (negatively), and the central galaxy's SFR (positively). The negative correlation with SMBH mass is shown for \texttt{IllustrisTNG} in the right panel of Figure~\ref{fig:fbar}, coloured by the average deviation in logarithmic SMBH mass relative to the median value. \citet{Davies2020} interpreted these correlations as an indication that early growth enables the SMBH to begin delivering efficient AGN feedback sooner, expelling more gas from the halo and extending the radiative cooling timescale of the remaining CGM. The reduced cooling efficiency inhibits replenishment of the ISM as it is consumed by star formation (or ejected by the associated stellar feedback), reducing the SFR or even quenching the galaxy. Similar trends have also recently been found in \texttt{SIMBA} \citep{Sorini2022}, and this qualitative consensus highlights that an intimate co-evolution of galaxies and their CGM is a major prediction of state-of-the-art simulations.

\section{The influence of environment}
\label{sec:environment}

Environment, in the context of galaxy evolution, can refer to a variety of important physical processes and effects. The cosmological environment governs the accretion rate onto haloes, which can be spatially inhomogeneous because of the filamentary structure of the cosmic web, as well as the rate of mergers. This profoundly impacts the growth of galaxies in combination with smaller scale processes, such as galactic winds. The large-scale structure and its inhomogeneity can lead to differences between cosmological and idealized simulations, though we will not further discuss this here.

The environment of a galaxy can also refer to whether or not it lives at the minimum of their host halo’s potential well, where the gas densities and cooling rates are highest. Satellite galaxies used to reside at the centres of their own haloes, but have since fallen into more massive haloes and are thus offset from the centre of the potential. Back-splash galaxies or flyby galaxies are those that were satellites in the past but have left the more massive halo they were previously in. They may therefore seem to not live in extreme environments, but can still have been strongly affected in the past. Note that there is no well-defined halo boundary and environmental effects driven by the halo can extend beyond the virial radius. When galaxies experience such effects before entering the more massive halo, this is also sometimes referred to as pre-processing.

\subsection{Satellite galaxies: Stripping \& Starvation} 
\label{sec:satellites}

Satellite galaxies make up a large proportion of the galaxy population and therefore must be included if we are to understand galaxy formation fully. Satellites make up more than 40 per cent of the population below a stellar mass of $3\times10^{10}$~M$_\odot$, though this decreases to less than 20 per cent above $M_\star=10^{11}$~M$_\odot$. Although we often think about processes involved in the formation of central galaxies and satellite galaxies separately, each satellite spent its early life (possibly most of its life) as a central and is therefore also shaped by the physical processes relevant for central galaxy formation. However, after becoming satellites, these galaxies evolve differently from central galaxies, because they experience various environmental effects, such as ram pressure and tidal stripping, as they move through the halo of their more massive companion. Note that a fraction of the galaxy population considered central galaxies are back-splash galaxies and have thus previously experienced stronger environmental effects.

Ram pressure is exerted on a body of gas when it moves through a medium with a different velocity. It scales with the density of the surrounding medium and the square of the velocity difference. Ram pressure only directly affects gas, whereas tidal effects are purely gravitational and thus affect all components of a satellite, i.e.\ dark matter, gas, and stars. When the ram pressure is high enough to remove the ISM of the satellite galaxy, this is typically referred to as ram pressure stripping \citep[e.g.][]{Gunn1972, Simpson2018}. With the ISM removed, the satellite can no longer form stars, so this results in relatively fast quenching of star formation. When only the gas in the halo of the satellite is removed, the galaxy can continue forming stars from its ISM until it is depleted. This is often referred to as starvation or strangulation and the quenching of star formation is much slower \citep[e.g.][]{vandeVoort2017, Wright2022}. In both cases, the satellite will not be able to replenish its ISM as it no longer has its own CGM acting as a reservoir from which it can accrete fresh gas. 

Modern cosmological simulations have sufficient resolution to capture these stripping processes, though resolution effects can make a quantitative difference, because galaxies that are less well resolved are more easily disrupted \citep[e.g.][]{Yun2019}. If simulations were able to fully model the multiphase ISM, it is likely that the cold, dense gas would be less easily stripped than in current simulation suites. \citet{Wright2022} find that starvation and ram pressure stripping of the ISM contribute a similar amount to the reduction of the SFR in satellite galaxies. Some evidence has been found that ram pressure stripping is stronger along the central galaxy's major axis than along its minor axis, pssibly because the density in the polar direction is reduced due to feedback-driven outflows \citep{MartinNavarro2021}.

Quenching, or the suppression of star formation, is a complicated and unsolved problem, as discussed before in Section~\ref{sec:colours}. Quenching in satellites depends on internal feedback processes as well as external stripping. Simulations have struggled to quantitatively reproduce observed quenched fractions at all stellar masses. There is general agreement that the majority of satellites are quenched, with the quenched fraction increasing towards lower masses \citep{Bahe2017, Donnari2021}. However, observations seem to show the opposite trend and very low quenched fractions for $10^9<M_\star<10^{10}\,\Msun$ \citep{DaviesL2019}. For an unbiased comparison, it will be necessary to process the simulations in the same way as is done for the observational measurements.

\subsection{Environmental effects: beyond halo mass}
\label{sec:env_beyond_mhalo}

The environment of a galaxy is dominated by properties of the dark matter-dominated halo it lives in \citep[e.g.][]{Crain2009}. Although straightforward to determine in simulations, halo mass is difficult to measure observationally. Instead, there are a variety of observational methods that can be used to quantify the environments of galaxies, such as distance to the N-th nearest neighbor. These often used environmental indicators correlate strongly with the halo mass \citep{Haas2012, Marasco2016}. Many of the differences identified based on environmental indicators exist because they are sensitive to host halo mass, and satellite galaxies behave differently from central galaxies, as discussed in Section~\ref{sec:satellites}.

In addition to a dependence of galaxy formation on halo mass, the larger scale environment the halo lives in can also influence the evolution of embedded galaxies to a certain extent. To study environmental effects that go beyond the halo mass of the system, it is important to remove the halo mass dependence and characterize the large-scale environment. This usually requires a large-scale structure finder, both in observations and simulations, or visual classification. \citet{RosasGuevara2022} build a void catalogue and compare galaxies living in those voids to those residing in the cosmic web. Overall the differences are fairly minor, e.g.\ average stellar masses can change by up to about 30 per cent. They find that this large-scale environment affects low-mass galaxies differently from high-mass galaxies, which likely indicates that there are a variety of processes at play. The galaxy mass in low-mass haloes is lower in voids, potentially due to a combination of lower accretion rates and lower merger rates, whereas it is higher in massive haloes, potentially because of lower black hole growth and lower amount of AGN feedback in void galaxies.

Ram-pressure stripping, as discussed in Section~\ref{sec:satellites} is not limited to only taking place inside galaxy haloes. The filaments of the cosmic web can also harbour strong accretion shocks and provide ram pressure, which can result in stripping of the gas in and around dwarf galaxies that pass through \citep{BenitezLlambay2013, Pasha2022, Herzog2022}. The ambient temperature of the gas in these filaments can be higher than the virial temperatures of dwarf galaxies, which means that they are not able to accrete fresh gas and thus remain gas-deficient. These galaxies have lower star formation rates and lower ISM masses than galaxies not affected by stripping in the cosmic web, yet they can appear quite isolated.

\section{Future outlook}
\label{sec:future}

The level of realism achieved by state-of-the-art cosmological hydrodynamical simulations of the galaxy population has advanced dramatically in the last decade. Despite this success, the outcomes are particularly sensitive to subgrid implementations of feedback processes. It is thus reasonable to argue that this success suffices only to establish that the basic sketch of galaxy formation theory within the $\Lambda$CDM cosmogony is plausible. The development of simulations that can be used to stress test a truly comprehensive theory of galaxy formation and evolution will require confrontation of the devil in the detail. Simulations of the galaxy population do not accurately reproduce the internal and vertical structure of galaxies and their reliance on subgrid methods to approximate the influence of unresolved physics leaves lingering degeneracies that diminish predictive power. Moreover, the simulations neglect physical processes known to be significant in certain regimes. From our suite of examples, only \texttt{IllustrisTNG} models magnetic fields, which could impact cosmic gas flows and thus affect the evolution of galaxies \citep{Pillepich2018b, vandeVoort2021}. The omission of magnetic fields also precludes realistic modelling of the influence of cosmic rays, which, owing to their ISM energy density being comparable with those of the thermal and magnetic pressures, likely influence galaxy-wide outflows \citep{Uhlig2012}. Similarly, thermal conduction influences the structure of the ICM and cooling onto massive galaxies \citep{Carilli2002} but is generally neglected. Modelling the UVB as a spatially-uniform radiation field that `switches on' at a fixed redshift is a particularly crude approximation for the evolution of galaxies during the EoR, when local sources of ionizing radiation can dominate the regulation of galaxy growth \citep{Wise2009,Trebitsch2017,Katz2020}. Locally-varying radiation fields, such as those produced by flickering AGN, can also overionise gas and lead to cooling rates being overestimated by orders of magnitude if equilibrium conditions are assumed \citep{Vasiliev2011,Richings2014a}.

It is natural to consider how the field will build on the foundation provided by the current state-of-the-art generation of hydrodynamic simulations of the galaxy population. There are several routes by which we envisage that progress will be made, beyond the usual pursuit of superior resolution \citep[which is needed to achieve converged results in the CGM, e.g.][]{vandeVoort2019}. Despite remarkable recent demonstrations of simulation codes running on hundreds of thousands of compute cores \citep[e.g.][]{Schaller2016,Pakmor2023}, subgrid models will remain a necessary component of simulations of the galaxy population for some years to come, precluding holistic, quantitative predictive power. There is, however, much scope to develop more detailed implementations, and to interface them with the numerical calculation on shorter spatial scales, enabling more detailed confrontations with observations and increasing the diversity of the lines of enquiry for which the simulations can offer authoritative predictions. Moreover, the ill-constrained parameters of the subgrid models used by leading simulations have been calibrated manually, by performing a small set of parameter-spanning simulations, analysing their outputs, and choosing updated parameters for the subsequent set based on the practitioner's intuition for the response of the galaxy population to the adjustment. Statistical methods exist to formalise this process and ensure that the plausible parameter space is efficiently explored. There is also much to be gained not only from improving the simulations themselves, but also the techniques used to analyse their outcomes.

The development and testing of new or improved treatments of physical processes, whether implemented numerically or subgrid, is usually pioneered using simulations that adopt idealised or zoomed cosmological initial conditions. At fixed resolution, such simulations are markedly less expensive than simulations of representative volumes or, for a fixed number of computational core hours, they allow galaxies to be evolved at much higher resolution. Such simulations have been used to pursue more detailed treatments of the multiphase ISM that require fewer assumptions, in particular relating to the star formation efficiency of dense gas  \citep[e.g.][]{Hopkins2014,Semenov2016,Kim2017}. \citet[][see also Fig.~\ref{fig:GSMF}]{Feldmann2023} showed recently that the use of these more detailed models in a periodic cosmological volume does not guarantee the emergence of realistic GSMF, even in the low stellar mass regime for which the included physical processes are expected to dominate galaxy regulation. A number of suites of zoom simulations focusing on dwarf galaxies have achieved sufficiently high resolution to model individual SNe explosions within a multiphase ISM \citep[e.g.][]{Wheeler2019, Agertz2020, Gutcke2022}.  Detailed study of these processes in idealized or zoom simulations paves the way towards development of coarse-grained descriptions that, although still approximate, can be more realistic than the simple (and diverse) treatments of stellar feedback used by today's galaxy population simulations. An extreme interpretation of this methodology is to treat outflows exclusively with a phenomenological approach, even on numerically-resolved scales \citep[e.g.][]{Huang2020}. Similarly, the multiphase structure of the CGM is not resolved in cosmological simulations, specifically the cool gas and the interfaces between the cool and hot CGM \citep[e.g.]{McCourt2018, Fielding2020}, see also Faucher-Gigu\`ere \& Oh, this volume. Accurately treating the cool and intermediate temperature gas may require developing subgrid models within the CGM.  

The recent discovery with JWST of multiple galaxies with spectroscopically-confirmed redshifts $z>10$ \citep{CurtisLake2023} highlights the urgent need to model the EoR galaxy population with explicit radiation hydrodynamics (RHD). Zoom simulations have proven useful for examination of the internal structure of individual galaxies in this regime \citep[e.g.][]{Pallottini2017} but suffer from potential selection biases and preclude examination of the influence of galaxies on the IGM as the latter undergoes a global phase transition. High-resolution RHD simulations of representative volumes remain extremely challenging, but \citet{Borrow2023} were able to show using simulations of small periodic volumes that the use of a spatially-uniform UVB fosters significant inaccuracies in the space density and internal properties of low-mass simulated galaxies in the EoR, and results in the density and temperature structure of the IGM being too uniform. 

In recent years, machine learning has steadily grown in influence in nearly every aspect of astrophysics \citep[see review by][]{Smith2023}. Machine learning techniques have emerged as an effective means of elucidating formally the complex relationships between the properties of simulated galaxies and those of their host dark matter haloes \citep{IcazaLizaola2021,Piotrowska2022}. The CAMELS (Cosmology and Astrophysics with Machine Learning Simulations) project \citep{VillaescusaNavarro2021} uses many simulations evolved with variations of the \texttt{IllustrisTNG} and \texttt{SIMBA} models, varying both cosmological and subgrid parameters, as a basis for the application of machine learning techniques. \citet{Jo2022} use the suite as a testbed for calibrating model parameters using neutral networks trained as emulators. Interestingly, they found that the emulators could identify parameter sets that accurately reproduce a GSMF drawn from the input simulations chosen as a `target observable', but none that reproduce a real observationally-inferred GSMF. It is unclear whether this signals a problem with the method, inconsistencies in the observationally-inferred GSMF, or a genuine limitation of the physics implemented within the input simulations. If the latter, this methodology may prove a particularly effective means of guiding the development of more sophisticated subgrid models. 

\section{Summary}
\label{sec:summary}

Cosmological, hydrodynamical simulations have catalysed significant strides in our understanding of the physics that governs the formation and evolution of galaxies. The much improved realism of the current generation of state-of-the-art simulations, relative to their predecessors, has engendered greater confidence in the insight obtained from the confrontation of simulations with observational data, and has diversified the lines of enquiry for which such comparisons are useful.

A crucial lesson learned from some of the first simulations of galaxies is that their evolution is strongly regulated by energetic feedback processes. Modern, realistic simulations model (at minimum) feedback associated with the formation and evolution of stellar populations (which dominates in low mass galaxies), and that associated with the accretion of gas onto SMBHs (which dominates in massive galaxies). However, it remains the case (and will do so for some years to come) that this modelling is achieved in an approximate fashion using subgrid models, whose governing parameters are ill-constrained. The greater realism of the current generation of simulations has therefore followed from the pragmatic approach of calibrating these parameters.

Simulations calibrated to reproduce the stellar masses and sizes of galaxies also reproduce many well-known observed scaling relations, e.g. the Tully-Fisher relation and the SFMS. This is often because they concern related properties, or because reproduction of the relation relies primarily on ensuring that galaxies of a fixed stellar mass are associated with dark matter haloes of the correct total mass and hence exhibit (for example) realistic space densities and cosmic matter inflow rates. Other scaling relations, particularly those related to the baryon cycle in haloes such as the mass -- metallicity relation, the baryon fraction -- halo mass relation, and galaxy -- absorber statistics, are more sensitive to the details of the adopted subgrid models. Quantitative agreement between the simulations and observations for these relations tends to require that the simulations be calibrated specifically to achieve it. The quantities in these relations also tend to be those for which numerical convergence is most challenging to achieve. Perhaps unsurprisingly then, there is a conspicuous absence of inter-simulation consensus in this regime.

Different subgrid prescriptions therefore appear able to reproduce many galaxy properties equally well, whilst yielding predictions for other properties (often those only weakly constrained by observations) that are markedly different. As we highlight in the summary boxes below, the current generation of state-of-the-art simulations of the galaxy population has therefore fostered important and enduring successes, but many outstanding challenges remain. The simulations provide a sound foundation from which to pursue the more sophisticated models needed to tackle these challenges and further our understanding of the galaxy population. This pursuit will of course also be driven by future observational discoveries, and more detailed characterisation of the CGM is likely to prove amongst the most fruitful avenues for constraining, in particular, the macroscopic effects of energetic feedback.

Finally, we remark that the public data release (with detailed accompanying documentation) of major simulation campaigns, while being a major undertaking for the simulation teams, has proven a tremendously successful exercise. It has enabled their use by many researchers who were not involved with the development of the simulations, and many astronomers without prior specialism in simulations, allowing more thorough exploitation and analysis of the simulations and more diverse comparisons with observations. We energetically encourage the developers of future simulation campaigns to follow suit.

% Summary Points
\begin{summary}[SUMMARY POINTS]
\begin{itemize}
\item By regulating star formation with plausible quantities of energetic feedback associated with the formation of stars and the growth of black holes, the current generation of state-of-the-art cosmological hydrodynamical simulations form a galaxy population with broadly realistic stellar masses and sizes.
\item Although the more realistic of these simulations were calibrated against present-day galaxy masses and sizes, the evolution of these quantities was not, and the observed evolutionary trends are also broadly reproduced.
\item Simulated galaxy populations exhibit the diversity of present-day morphologies exemplified by the Hubble Sequence, as a natural outcome of the diversity of galaxy assembly histories and the intrinsic properties of their host haloes. 
\item The observed clustering of galaxies, as a function of stellar mass, galaxy colour, and atomic gas content, is reproduced by simulations on the scales they are able to reliably sample and adequately resolve.
\item The simulations reproduce many key galaxy scaling relations, elucidating their origin. A key example is the stellar -- halo mass relation, which has been shown to emerge primarily in response to gas expulsion in the low-mass regime, and throttling of cooling from the CGM onto the ISM in more massive haloes, due to AGN feedback. 
\item The simulations also illuminate the origin of scatter about scaling relations, with key examples being the scatter in star formation rate, and in metallicity, at fixed stellar mass. Both can be explained in terms of the balance of gas flows into and out of galaxies, star formation, and black hole growth. 
\item Once armed with a realistic model, examination of partner simulations in which model components are adjusted or toggled has proven an effective approach to illuminating the sensitivity of galaxy properties and observables to physical processes. 
\end{itemize}
\end{summary}

% Future Issues
\begin{issues}[FUTURE ISSUES]
\begin{itemize}
\item Simulations of the galaxy population still rely on simplified subgrid models to treat unresolved physical processes, including feedback associated with the formation of stars and the growth of black holes. Simulations thus remain distantly removed from ab initio theory.
\item Physical processes treated with subgrid models dominate the systematic uncertainty on the properties of simulated galaxies. Since different subgrid models can produce broadly realistic simulated galaxy populations, significant degeneracies remain between state-of-the-art simulation suites and predictive power is limited.
\item Even in simulations whose subgrid models were designed to minimise their sensitivity to resolution, the convergence behaviour of some physical properties (particularly those related to dense gas) can be poor, eroding confidence in outcomes.
\item The detail of simulations of the galaxy population remains relatively poor: the internal and vertical structure of galaxy discs is unrealistic owing to the simplistic subgrid modelling of interstellar gas.
\item There is relatively little quantitative inter-simulation consensus concerning the properties of gas flows and the CGM, which are particularly sensitive to the implementation of feedback processes. 
\end{itemize}
\end{issues}

%Disclosure
\section*{Disclosure statement}
The authors are not aware of any affiliations, memberships, funding, or financial holdings that might be perceived as affecting the objectivity of this review. 

% Acknowledgements
\section*{Acknowledgements}
We extend our gratitude to the many colleagues and collaborators with whom we have enjoyed illuminating discussions that have helped to shape this review. We thank Romeel Dav\'e, Jon Davies, Alex Hill, R\"udiger Pakmor, and Michael Tremmel for providing data used in the figures, and the \texttt{EAGLE} and \texttt{IllustrisTNG} teams for making their data publicly accessible.  We are particularly grateful for Debora Sijacki's contribution to the early phase of assembling this review, and Tim Davis, Matthieu Schaller, and Renske Smit for valuable discussions. We thank Roselyn Lowe-Webb and Luis Ho, respectively our production and scientific editors at ARA\&A, for their patience and encouragement, and Glenda Mahoney for assistance with figures, in particular for creating Figure~\ref{fig:scales}. The authors are supported by Royal Society University Research Fellowships. 

% References
%
% Margin notes within bibliography
%\section*{LITERATURE\ CITED}
\bibliographystyle{ar-style2}
\bibliography{bibliography}

\end{document}